\documentclass[pra,reprint,superscriptaddress,bibnotes]{revtex4-1}

\usepackage[T1]{fontenc}
\usepackage[utf8]{inputenc}
\usepackage[australian]{babel}
\usepackage[a4paper,centering,hmargin=1.75cm,vmargin=2cm]{geometry}
\usepackage{amsmath,amssymb,bm}
\usepackage{graphicx,xcolor}
\usepackage{titlesec}
\usepackage{microtype}
\usepackage{braket,siunitx}
\usepackage{bbold}

\newcommand{\hc}{\mathrm{h.c.}}

\begin{document}

\title{Analog quantum simulation of chemical dynamics}

\affiliation{School of Chemistry, University of Sydney, NSW 2006, Australia}
\affiliation{School of Physics, University of Sydney, NSW 2006, Australia}
\affiliation{ARC Centre of Excellence for Engineered Quantum Systems, University of Sydney, NSW 2006, Australia}
\affiliation{University of Sydney Nano Institute, University of Sydney, NSW 2006, Australia}

\author{Ryan J. MacDonell}
\affiliation{School of Chemistry, University of Sydney, NSW 2006, Australia}
\affiliation{University of Sydney Nano Institute, University of Sydney, NSW 2006, Australia}

\author{Claire E. Dickerson}
\affiliation{School of Chemistry, University of Sydney, NSW 2006, Australia}
\affiliation{School of Physics, University of Sydney, NSW 2006, Australia}
\affiliation{ARC Centre of Excellence for Engineered Quantum Systems, University of Sydney, NSW 2006, Australia}
\affiliation{University of Sydney Nano Institute, University of Sydney, NSW 2006, Australia}

\author{Clare J.T. Birch}
\affiliation{School of Chemistry, University of Sydney, NSW 2006, Australia}
\affiliation{University of Sydney Nano Institute, University of Sydney, NSW 2006, Australia}

\author{Alok Kumar}
\affiliation{Department of Chemistry, Indian Institute of Technology Bombay, Mumbai, 400076, India}

\author{Claire~L.~Edmunds}
\affiliation{School of Physics, University of Sydney, NSW 2006, Australia}
\affiliation{ARC Centre of Excellence for Engineered Quantum Systems, University of Sydney, NSW 2006, Australia}
\affiliation{University of Sydney Nano Institute, University of Sydney, NSW 2006, Australia}

\author{Michael J. Biercuk}
\affiliation{School of Physics, University of Sydney, NSW 2006, Australia}
\affiliation{ARC Centre of Excellence for Engineered Quantum Systems, University of Sydney, NSW 2006, Australia}
\affiliation{University of Sydney Nano Institute, University of Sydney, NSW 2006, Australia}

\author{Cornelius Hempel}
\affiliation{School of Physics, University of Sydney, NSW 2006, Australia}
\affiliation{ARC Centre of Excellence for Engineered Quantum Systems, University of Sydney, NSW 2006, Australia}
\affiliation{University of Sydney Nano Institute, University of Sydney, NSW 2006, Australia}

\author{Ivan Kassal}
\email{ivan.kassal@sydney.edu.au}
\affiliation{School of Chemistry, University of Sydney, NSW 2006, Australia}
\affiliation{University of Sydney Nano Institute, University of Sydney, NSW 2006, Australia}

\begin{abstract}
Ultrafast chemical reactions are difficult to simulate because they involve entangled, many-body wavefunctions whose computational complexity grows rapidly with molecular size.
In photochemistry, the breakdown of the Born-Oppenheimer approximation further complicates the problem by entangling nuclear and electronic degrees of freedom.
Here, we show that analog quantum simulators can efficiently simulate molecular dynamics using commonly available bosonic modes to represent molecular vibrations. Our approach can be implemented in any device with a qudit controllably coupled to bosonic oscillators and
with quantum hardware resources that scale linearly with molecular size, and offers significant resource savings compared to digital quantum simulation algorithms.
Advantages of our approach include a time resolution orders of magnitude better than  ultrafast spectroscopy, the ability to simulate large molecules with limited hardware using a Suzuki-Trotter expansion, and the ability to implement realistic system-bath interactions with only one additional interaction per mode.
Our approach can be implemented with current technology; e.g., the conical intersection in pyrazine can be simulated using a single trapped ion. Therefore, we expect our method will enable classically intractable chemical dynamics simulations in the near term.
\end{abstract}

\maketitle

\section{Introduction}
Computational chemistry aims to predict energies, structures, reactivity, and other properties of molecules. Although molecular dynamics is, in principle, best
simulated with a fully quantum-mechanical treatment of coupled electrons and nuclei, the computational cost of
doing so scales exponentially with molecular size, making it intractable for most chemical systems. The cornerstone of
the vast majority of quantum-chemistry methods is the Born-Oppenheimer (adiabatic) approximation, which neglects the coupling between electronic and nuclear degrees of freedom~\cite{domcke04}. This approximation fails in regions of strong non-adiabatic coupling, particularly near
degeneracies between electronic states, known as conical intersections~\cite{domcke04}. Non-adiabatic couplings are essential to photochemistry, where conical
intersections act as funnels from one electronic state to another on ultrafast timescales (those comparable to nuclear vibrational periods, fs--ps)~\cite{domcke04,cederbaum81}.
State-of-the-art algorithms, such as (multilayer) multiconfigurational
time-dependent Hartree (MCTDH)~\cite{worth08,wang15}, can significantly reduce the simulation cost using a careful
choice of basis-set contractions optimized for each system, enabling simulations of particular systems with tens to hundreds of modes~\cite{meng13,xie15,schulze16}.
Nevertheless, the optimal form of MCTDH wavefunctions cannot be predicted a priori, and the method---like all fully quantum-mechanical treatments of chemical dynamics---has an exponential worst-case scaling with system size.

Quantum computing promises to reduce the steep computational cost of simulating quantum systems by representing the problem on a controllable quantum
device~\cite{Feynman.1982,lloyd95,nielsen10,Buluta2009,blatt2012,aspuru2012,bloch2012,RevModPhys.86.153,schafer2020}. Most chemical applications of quantum computing have focused on time-independent observables~\cite{aspuruguzik05,lanyon10,yung14,peruzzo14,li17,Kandala.2017,colless18,hempel18,sawaya19,nam20,rubin20}, in an effort
to reduce the cost of electronic-structure methods. Such quantum methods optimize trial wavefunctions and map observables onto registers of
qubits, and can predict molecular properties with linearly many qubits and polynomially many quantum gates as a function of molecule size~\cite{aspuruguzik05,lanyon10}.
Due to limitations in qubit count and coherence times of current quantum computers, many recent methods use hybrid approaches, such as variational quantum
eigensolvers, to divide the calculation into classical and quantum tasks~\cite{peruzzo14,li17,Kandala.2017,colless18,hempel18,nam20}.
By contrast, few quantum algorithms focus on the simulation of molecular dynamics. Proposed methods include the simulation of electronic~\cite{li17} and vibrational~\cite{mcardle19,sawaya20}
dynamics and the coupling between them~\cite{kassal08,ollitrault20b}. Although each of these methods scales polynomially with system size, their qubit
requirements restrict them to small model systems.

Analog quantum simulators provide an alternative approach to quantum simulation by mapping a desired Hamiltonian onto a purpose-built quantum system~\cite{Buluta2009,blatt2012,aspuru2012,bloch2012,RevModPhys.86.153}. As a result, the dynamics of the simulator directly corresponds to the dynamics of the simulated system, and allowing any observables to be read out, in principle, using suitable measurements.
Qubit-based analog simulators have already been used to demonstrate quantum advantage by carrying out classically-intractable simulations of spin models~\cite{britton12,zhang17}. As a chemical application, ultracold atoms in optical lattices were proposed to simulate the electronic degrees of freedom of a molecule in a grid-based adiabatic picture~\cite{arguelloluengo19}.

Particularly promising recent approaches to analog quantum simulation have taken advantage of bosonic modes natively present in certain architectures, including photonic chips, microwave resonators, and trapped ions.
Using native bosonic degrees of freedom gives analog simulators an advantage over digital ones, which would require many qubits to accurately represent a single harmonic oscillator.
Chemical applications of bosonic simulation have mainly focused on the vibrational degrees of freedom of molecules, such as vibrational
dynamics~\cite{sparrow18,chen21} and Franck-Condon vibronic spectra~\cite{yung14,huh15,wang20,jnane20,chen21}. Approaches that include the entanglement of internal states to bosonic modes could simulate more complex processes, such as the dynamics of fermionic
lattice models~\cite{casanova12,lamata14}, with possible extensions to electron-nuclear dynamics for quantum chemistry~\cite{lamata14,lamata18}, as well as models of charge and energy transfer~\cite{garciaalvarez15,gorman18,lemmer18,schlawin20}.

Here, we show that analog quantum simulators can efficiently simulate non-adiabatic chemical dynamics.
Our approach can be implemented using any quantum system containing a qudit with controllable couplings to a set of bosonic modes, a device we call a mixed qudit-boson (MQB) simulator.
We achieve linear scaling in molecular size by mapping a molecule with $d$ electronic states and $N$ vibrational modes onto an MQB with a $d$-level qudit and $N$ bosonic modes.
We analyze the non-adiabatic dynamics of pyrazine as an example, showing it can be simulated using a single trapped-ion MQB, and we detail how to extend the procedure to large systems using a Trotterization scheme and to open systems using laser cooling.

\section{Hamiltonian mapping}
\begin{figure}[t]
    \centering
    \includegraphics[width=0.95\columnwidth]{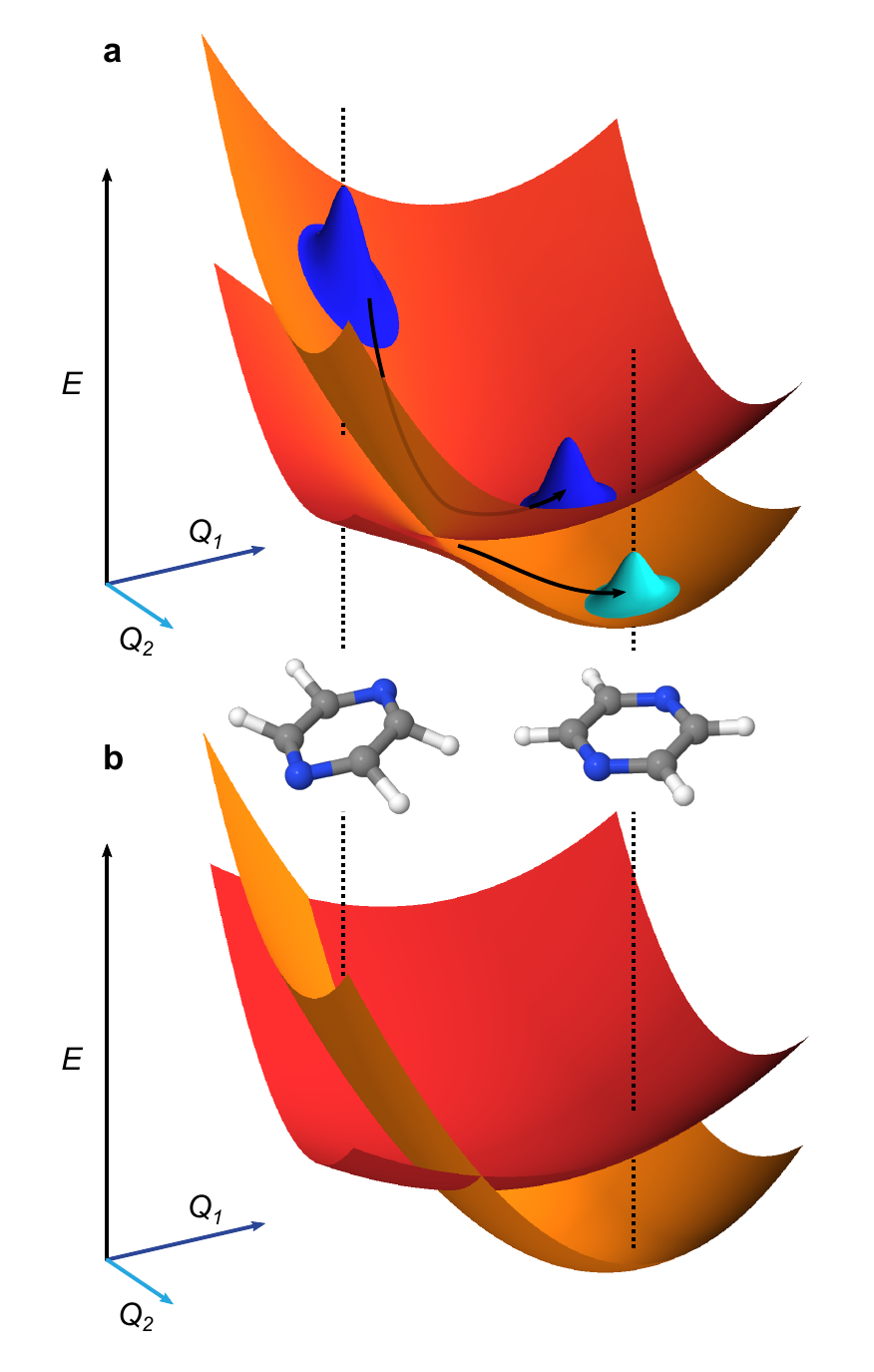}
    \caption{
    Representations of molecular potential-energy surfaces and their link to chemical dynamics.
    (\textbf{a})~Adiabatic surfaces correspond to the eigenvalues of the Born-Oppenheimer Hamiltonian at all nuclear displacements. The upper and lower surfaces shown correspond to higher- and lower-energy electronic states. Two inset molecules illustrate different molecular geometries at two sets of nuclear coordinates $Q_1$ and $Q_2$. An initial wavepacket (blue, top left) can slide down the upper surface and, upon reaching the conical intersection (the cusp where the two surfaces are degenerate) it can split into two entangled branches (blue and cyan), one on each electronic surface.
    (\textbf{b})~Diabatic surfaces (red, orange) of a vibronic coupling Hamiltonian (Eq.~\ref{eq:vcham}) formed from the adiabatic surfaces in~\textbf{a}. The diabatic surfaces are analytic, avoiding numerical divergences near conical intersections. The couplings between the diabatic surfaces are not shown.
    }
    \label{fig:surfs}
\end{figure}

Our approach uses vibronic coupling (VC) Hamiltonians, which are widely used to model chemical dynamics. They express molecular electronic matrix elements as analytical functions of the nuclear coordinates, often as a power series~\cite{domcke04},
\begin{align}
    \hat{H}_\mathrm{VC} &= \frac{1}{2}\sum_j \omega_j(\hat{Q}_j^2 + \hat{P}_j^2) + \sum_{n,m} \hat{C}_{n,m}\ket{n}\bra{m}, \label{eq:vcham}\\
    \hat{C}_{n,m} &= c_0^{(n,m)} + \sum_j c_j^{(n,m)} \hat{Q}_j + \sum_{j,k} c_{j,k}^{(n,m)} \hat{Q}_j \hat{Q}_k + \cdots, \label{eq:vcpoly}
\end{align}
where $\ket{n}$ is the $n$th electronic state and $\hat{Q}_j = (\mu_j \omega_j)^{1/2} \hat{q}_j$ and $\hat{P}_j = (\mu_j \omega_j)^{-1/2} \hat{p}_j$
are the dimensionless position and momentum of mode $j$ (which has reduced mass $\mu_j$, frequency $\omega_j$, position $\hat{q}_j$, and momentum
$\hat{p}_j$). We set $\hbar = 1$ throughout. The coefficients  $c_0^{(n,m)}$, $c_j^{(n,m)}$, etc. are the (real) expansion coefficients of the molecular potential energy about the reference
geometry, typically the minimum of the ground electronic state.
While adiabatic potential energy surfaces obtained from electronic structure methods are ill-behaved,
with features such as conical intersections and singular kinetic-energy couplings, they can be transformed to the diabatic surfaces and couplings of VC
Hamiltonians given by Eq.~\ref{eq:vcpoly} (Fig.~\ref{fig:surfs})~\cite{domcke04}.
For most applications, linear vibronic coupling (LVC) or quadratic vibronic coupling (QVC) models accurately represent photochemical observables~\cite{domcke04,seidner92}. However,  the expansion order of the Hamiltonian has little effect on the classical memory required to store the wavefunction.

\begin{figure*}
    \centering
    \includegraphics[width=0.92\textwidth]{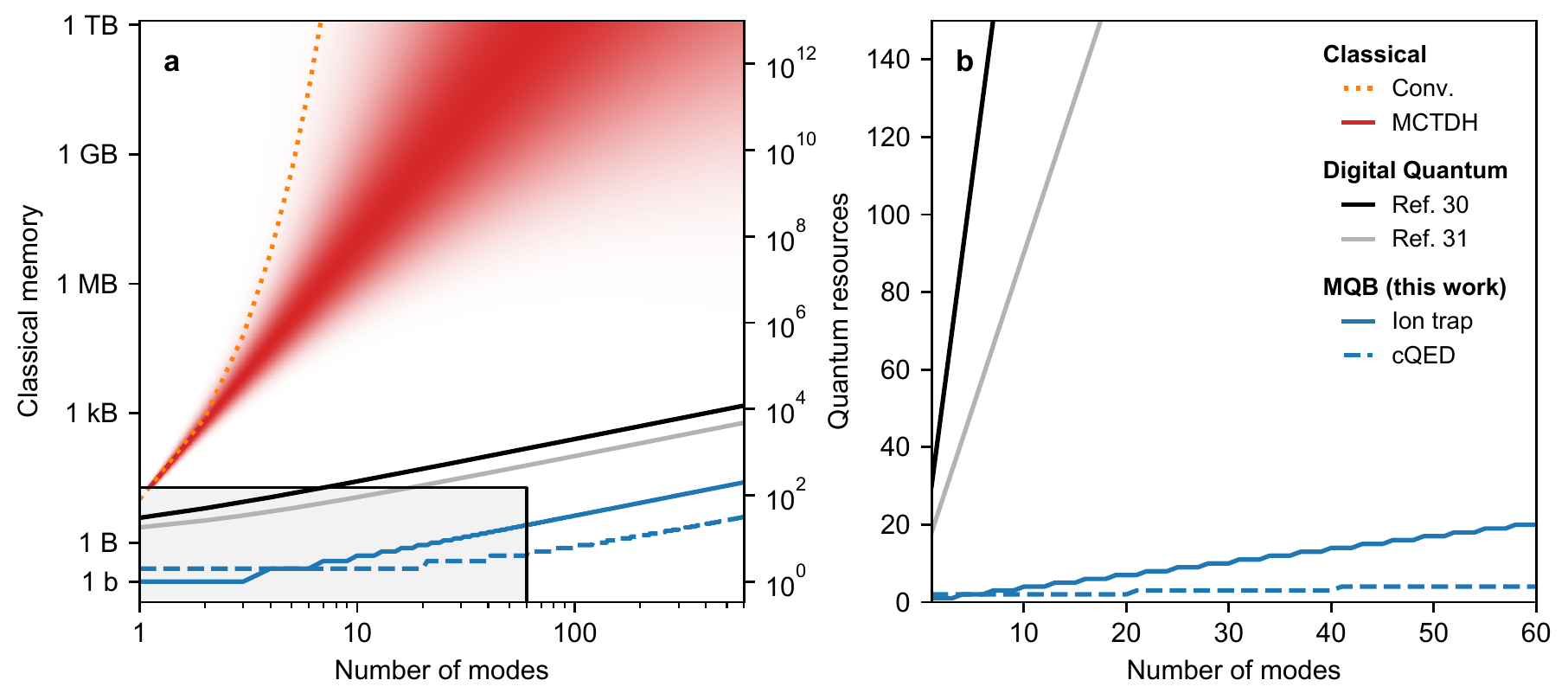}
    \caption{Resource (space) scaling of different methods for simulating vibronic dynamics.
    (\textbf{a}) Classical memory required for conventional (Conv.) and MCTDH approaches in bits (b) and bytes ($\mathrm{B = 8b}$), and quantum hardware resources for digital and MQB approaches. For MCTDH, an approximate range is shown, because the performance of the method depends on the Hamiltonian and cannot be accurately predicted a priori. The quantum resources are qubits (for digital algorithms, assuming 8~qubits per simulated mode), ions (for ion-trap MQB, assuming 3~vibrations per ion), and microwave resonators (for cQED MQB, assuming 20~modes per resonator).
    (\textbf{b}) Linear-scale comparison of quantum approaches (range as the grey box in \textbf{a}).}
    \label{fig:scale}
\end{figure*}

To map VC Hamiltonians onto MQB simulators, we exploit the similarities between terms in $\hat{H}_\mathrm{VC}$ and common entangling gates in MQB architectures.
We start by expressing an LVC model in the interaction picture,
\begin{multline} \label{eq:inter}
    \hat{H}_I^{\mathrm{LVC}} = \sum_{n, m}^d \sum_{j=1}^N \frac{c_j^{(n,m)}}{\sqrt{2}} \ket{n}\bra{m} (\hat{a}^\dag_j e^{i\omega_j t} + \hc )\\
    + \sum_n c_0^{(n,n)} \ket{n}\bra{n},
\end{multline}
where $\hat{a}^\dag_j$ and $\hat{a}_j$ are the bosonic creation and annihilation operators such that $\hat{Q}_j = (\hat{a}^\dag_j+ \hat{a}_j)/\sqrt{2}$ and $c_0^{(n,n)}$ is the potential energy of state $n$ at the reference geometry.
$\hat{H}_I^{\mathrm{LVC}}$ is a generalized Jaynes-Cummings Hamiltonian of a $d$-level system coupled to $N$ bosonic modes,
the same interaction that is used for digital quantum computation with existing architectures, including ion traps and cQED.
Typically, ion-trap quantum computers encode qubits in the electronic states of the ions, and lasers are used to entangle them by transient coupling through the ions' vibrations in the trapping potential~\cite{molmer99,leibfried03,lee05}. Likewise, circuit quantum electrodynamics (cQED) uses microwave cavities coupled to superconducting qubits to implement multi-qubit gates~\cite{blais04}.
Any MQB architecture can be described, to first order, with the Hamiltonian
\begin{align} \label{eq:mqb}
    \hat{H}_I^{\mathrm{MQB}} &= \sum_{n} \sum_j \Theta_{n,j}' \ket{n}\bra{n} (\hat{a}^\dag_j e^{i\delta_j t} + \hc) \nonumber\\
    &+ \sum_{n \neq m} \sum_k \Omega_{n,m,k}' \ket{n}\bra{m}(\hat{a}^\dag_k e^{i\delta_k t} + \hc) \nonumber\\
    &+ \frac{1}{2}\sum_n \chi_n \ket{n}\bra{n}.
\end{align}
Each of the parameters in $\hat{H}_I^{\mathrm{MQB}}$ can be tuned using well-developed light-matter interactions.
For example, for an ion trap, $\Theta_{n,j}'$ corresponds to a time-dependent AC Stark shift from a pair of non-copropagating lasers, $\Omega_{n,m,k}'$ is proportional to the
Rabi frequency of a M{\o}lmer-S{\o}rensen interaction, $\chi_n$ is a time-independent AC Stark shift and $\delta_j$ is the detuning from (sideband) resonance for mode $j$ (see Appendix~A for details).
Therefore, because Eq.~\ref{eq:mqb} is of the same form as Eq.~\ref{eq:inter}, the terms of the MQB Hamiltonian can be tuned to match those of an LVC Hamiltonian.

The MQB approach can also be extended to higher-order expansions; the second-order terms of a QVC Hamiltonian, in the interaction picture, become
\begin{multline}\label{eq:intqvc}
    \hat{H}_I^{\mathrm{QVC}} = \sum_{n,m}^d \sum_{j,k}^N \frac{c_{j,k}^{(n,m)}}{2} \ket{n}\bra{m}
    \hat{a}^\dag_j \hat{a}_k e^{i(\omega_j - \omega_k) t}.
\end{multline}
Both the dispersive ($j = k$) and mode-mixing ($j \neq k$) terms have been demonstrated experimentally in trapped ions~\cite{pedernales15,marshall16} and cQED~\cite{blais20,gao18}.

Overall, the numbers of qudit states and bosonic modes required for the MQB mapping scale linearly with the corresponding numbers of molecular degrees of freedom, a significant improvement over the exponential classical scaling of fully quantum-mechanical chemical dynamics methods.
Fig.~\ref{fig:scale} shows that even optimised classical methods such as MCTDH are limited to tens of modes for many systems, meaning that small to medium MQBs could achieve quantum advantage.
Fig.~\ref{fig:scale} also shows that the MQB approach can simulate chemical dynamics with a significant improvement in quantum hardware requirements compared to digital approaches~\cite{kassal08,ollitrault20b}.
Our advantage comes from the larger Hilbert space in a bosonic mode, which would otherwise require many qubits to simulate, possibly with complicated interactions between them being required to simulate simple bosonic processes.
In MQB architectures, this advantage can be decisive, because a single well-controlled bosonic mode might be comparably difficult to implement experimentally as a single well-controlled qubit.

The advantage of the MQB approach is due to the natural mapping between molecular vibrations and MQB bosonic modes, which obviates the need, present on digital computers, to represent the large bosonic Hilbert space using qubits.
The precise quantum requirements depend on the nature of the MQB simulator: if trapped ions are used, each ion provides three vibrational modes, while in cQED, each resonator could yield dozens of microwave modes~\cite{sundaresan15,kollar19}.
The cost of the simulation can also be measured in the total number of interaction terms required, which scales as $N^k d (d + 1)/2$ for an order-$k$ expansion of Eq.~\ref{eq:vcpoly}.
The vast majority of VC Hamiltonians are LVC models, which scale linearly with the number of vibrational modes, or QVC models, which scale quadratically~\cite{domcke04}.
The number of vibrations is a measure of the size of the molecule because it equals $N = 3N_n - 6$ in a molecule with $N_n$ atoms.
Furthermore, the simulation may be simplified by excluding certain modes or interactions that do not participate in the dynamics, either exactly (due to symmetry) or approximately (due to weak coupling).

\begin{figure*}
    \centering
    \includegraphics[width=0.7\textwidth]{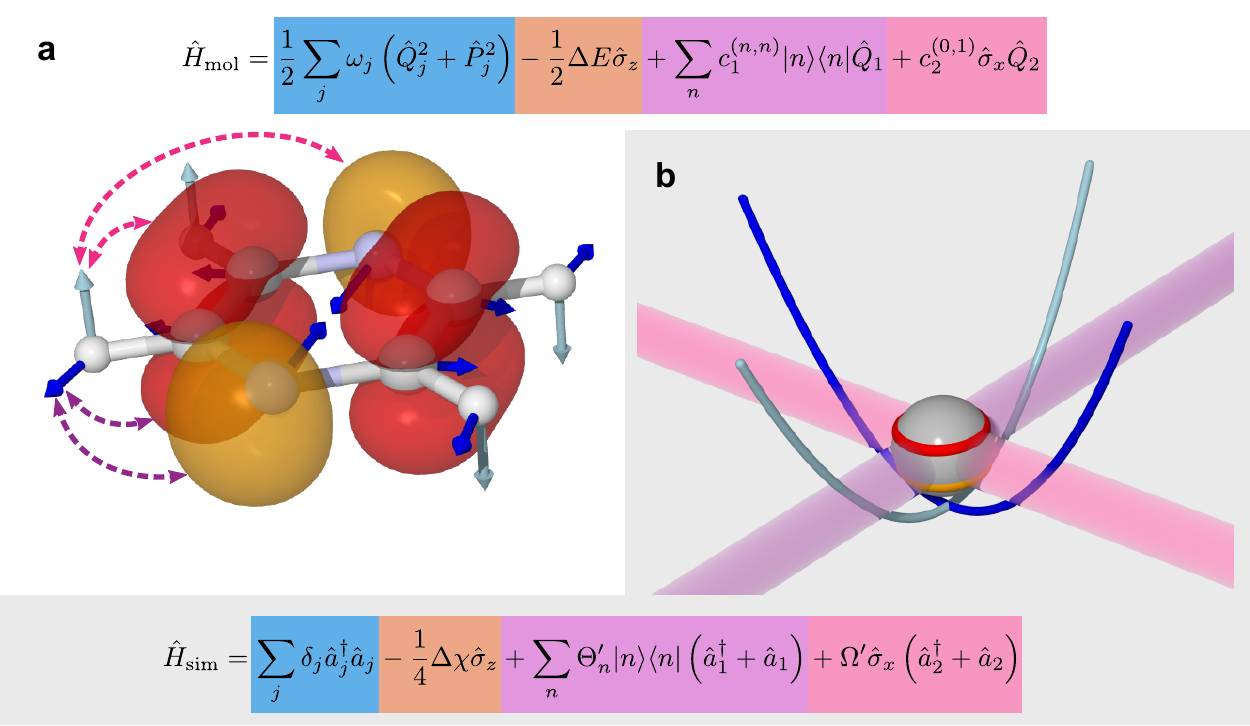}
    \caption{Example of the mapping between the degrees of freedom of a molecule (pyrazine, Eq.~\ref{eq:pyrham}) and an MQB simulator (one trapped-ion qubit, Eq.~\ref{eq:2dion}). (\textbf{a})~The degrees of freedom of the molecule: electronic, shown as red and orange electronic orbitals; vibrational, shown as cyan and blue atomic displacements;
    and their couplings shown in purple and magenta.
    (\textbf{b})~The corresponding degrees of freedom for the trapped ion: qubit states (red and orange levels); bosonic modes (cyan and blue potential curves); and laser-mediated interactions (purple and magenta). }
    \label{fig:map}
\end{figure*}

\section{Analog MQB simulation}
An MQB simulation of molecular dynamics comprises three steps: initialization, evolution, and measurement.

Many initial states can readily be prepared on an MQB simulator using established techniques. In particular, in most photochemical studies, the Franck-Condon principle implies that photoexcitation only promotes the electronic degree of freedom to an excited state, leaving the vibrations unchanged~\cite{domcke04}. On the MQB simulator, this corresponds to a change of the qudit state with no change
to the bosonic state. In more general cases, superpositions of simulator qudits states could be prepared to simulate molecules with nearly degenerate electronic states or undergoing broadband excitation.
Sometimes, it is convenient to work with displaced nuclear coordinates, in which case the initial state must be displaced as well, as discussed in Appendix~B.

Simulating the time evolution involves implementing qudit-boson interactions according to Eqs.~\ref{eq:inter} and \ref{eq:intqvc} to mimic the VC Hamiltonian for the duration of the evolution.
Given the correct light-matter interactions, the MQB simulator evolves in real time with the same dynamics as
the VC model, except that the Hamiltonian is scaled by a factor $F$ so that the evolution occurs on the natural
timescale of the MQB simulator. The scaling freedom is a major advantage for the simulation: whereas molecular vibronic frequencies are typically 10-100 THz, MQB simulators operate
at frequencies that are many orders of magnitude less (e.g., kHz for trapped ions, $F \sim 10^{-9}$, and GHz for cQED, $F \sim 10^{-3}$). The dynamics on the simulator thus occur in extreme slow motion relative to the simulated molecule, meaning the quantum simulator, like classical algorithms, has a much greater time resolution than ultrafast molecular spectroscopy.

Finally, after evolution for the desired amount of time, the MQB simulator can be probed to measure observables.
Most chemically important observables do not require knowledge of the full wavefunction (which would require exponentially scaling full state tomography), and are instead easily measurable properties of the qudits or the bosons, such as the electronic-state populations or nuclear positions and momenta. Most MQB architectures have established
methods for measuring qudit states; boson observables can also be measured, often by first mapping them onto the qudits~\cite{wang20,gerritsma11}.
Due to the statistical nature of quantum measurement, observables must be averaged over many experiments, while repeating the experiment for different simulation times would yield temporal information. Importantly, the results at any
simulation time are independent of the measurements from previous times and the time intervals at which the measurements are made.

\section{Example: 2D LVC model of pyrazine}

To demonstrate our approach, we consider the MQB simulation of a two-state, two-mode (or two-dimensional, 2D) LVC model of pyrazine using a single trapped ion (Fig.~\ref{fig:map}).
Pyrazine is a canonical model system for molecular dynamics because the conical
intersection between its two excited electronic states, the optically bright $\pi\pi$* state and the dark $n\pi$* state, can be described with only two modes. The vibrations
comprise a fully symmetric (tuning) mode and a symmetry-breaking (coupling) mode~\cite{seidner92,kuhl02}. The ground electronic state is energetically separated from and weakly coupled
to the excited states, meaning it can be excluded from the dynamics simulation~\cite{seidner92}. The 2D LVC Hamiltonian is
\begin{multline} \label{eq:pyrham}
    \hat{H}_\mathrm{mol} = \frac{1}{2}\sum_{j=1}^2\omega_j(\hat{Q}_j^2 + \hat{P}_j^2) - \frac{1}{2}\Delta E \hat{\sigma}_z \\
    + \sum_{n=0}^1 c_1^{(n,n)} \ket{n}\bra{n} \hat{Q}_1 + c_2^{(0,1)} \hat{\sigma}_x \hat{Q}_2,
\end{multline}
where $\ket{0}$ and $\ket{1}$ are the n$\pi$* and $\pi\pi$* states, modes 1 and 2 are the tuning and coupling modes, $\hat{\sigma}_z = \ket{0}\bra{0} - \ket{1}\bra{1}$ and
$\hat{\sigma}_x = \ket{0}\bra{1} + \ket{1}\bra{0}$ are Pauli matrices, and $\Delta E = c_0^{(1,1)} - c_0^{(0,0)}$ is the electronic energy difference at the ground-state-minimum geometry~\cite{seidner92}. The initial state of the simulation, corresponding to Franck-Condon excitation, is in the ground vibrational state, with the electronic state promoted to the $\pi\pi$* state, i.e., $\ket{1} \otimes \ket{v_1 = 0} \otimes \ket{v_2 = 0}$ for vibrational eigenstates $v_j$.

\begin{figure*}[t]
    \centering
    \includegraphics[width=\textwidth]{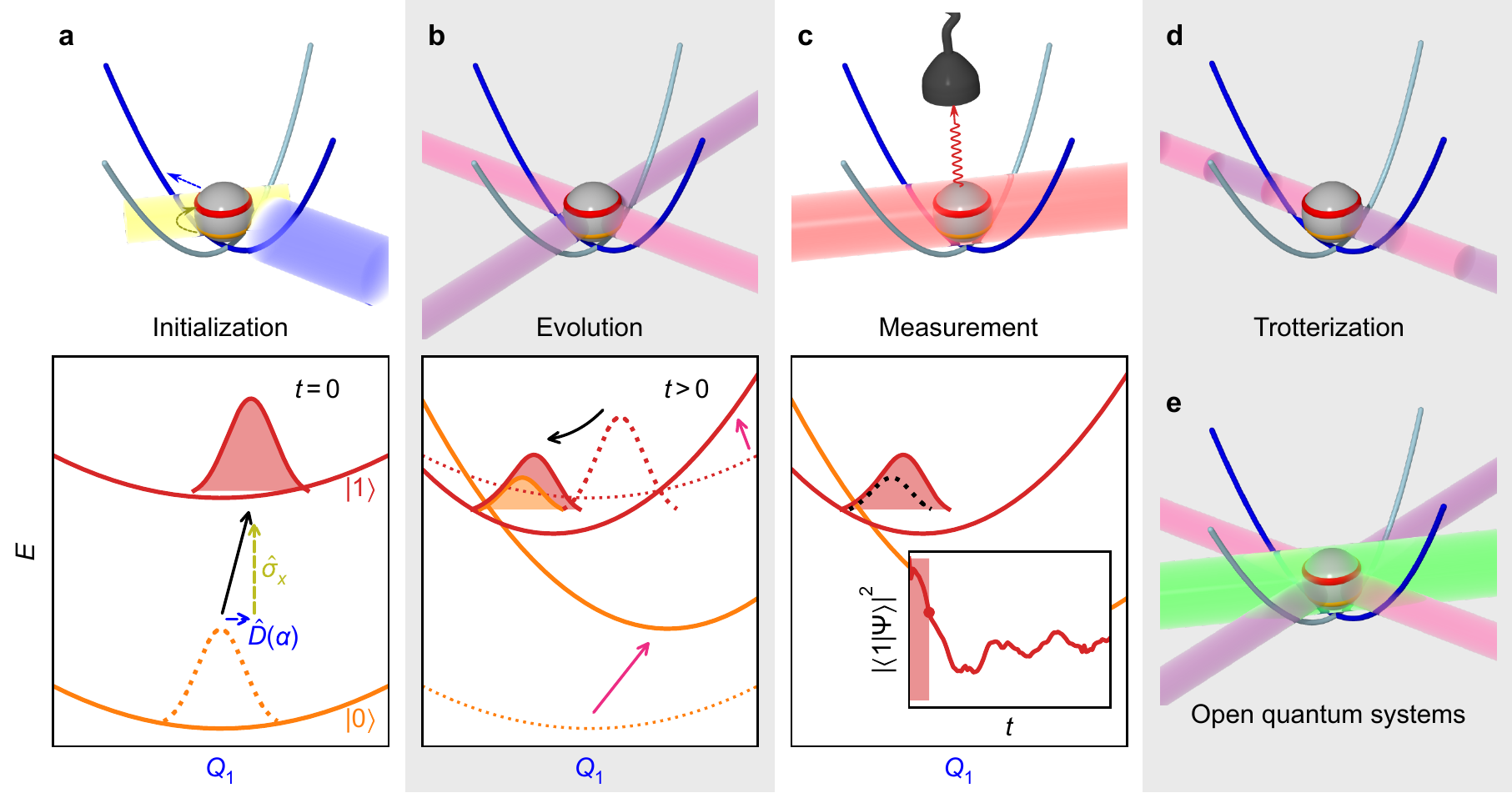}
    \caption{Illustrative molecular-dynamics simulation using an MQB simulator (trapped-ion qubit for concreteness).
    (\textbf{a})~The initial wavefunction is prepared by an excitation of the qudit state (yellow arrow/laser) and a displacement along the relevant mode(s) (blue arrow/laser).
    (\textbf{b})~The VC Hamiltonian is implemented through interactions with laser fields (purple and magenta, see Eq.~\ref{eq:mqb}),
    which are equivalent to displacements and couplings between potential energy surfaces. These interactions are maintained for the duration of the simulation.
    (\textbf{c})~After evolution by the effective
    Hamiltonian, observables of the wavefunction (inset: excited state population) are measured by a detector (e.g., fluorescence of the qubit state induced by the
    red laser). Multiple measurements are required to obtain expectation values, and the time dependence of observables is obtained by varying the time delay from initialization to measurement.
    (\textbf{d})~The evolution step
    can be discretized using a Suzuki-Trotter expansion into a series of short laser interactions, particularly for the treatment of many modes.
    (\textbf{e})~Open quantum systems can be simulated by adding sideband laser cooling (green) during the evolution step.}
    \label{fig:scheme}
\end{figure*}

In the absence of a laser field, the Hamiltonian of a single, two-level trapped ion is
\begin{align}
    \hat{H}_\mathrm{ion} &= \sum_{j=1}^2 \omega_j^\mathrm{ion} \hat{a}_j^\dag\hat{a}_j - \frac{1}{2}\omega_0\hat{\sigma}_z, \label{eq:ion}
\end{align}
where $\omega_0$ is the frequency difference between qubit states and $\omega_j^\mathrm{ion}$ is the trap frequency of mode $j$. The diagonal and off-diagonal couplings of Eq.~\ref{eq:pyrham} can be generated, respectively, by Raman interactions with bichromatic laser fields having frequency differences $\omega_j^\mathrm{ion} - \delta_j$ and $\omega_0 \pm (\omega_j^\mathrm{ion} - \delta_j)$, where $\delta_j$ is a detuning~\cite{lee05}.
The resulting Hamiltonian can be expressed in a rotating frame (i.e., by transforming Eq.~\ref{eq:mqb}) into the time-independent form
\begin{multline}
    \hat{H}_\mathrm{sim} = \sum_{j=1}^2 \delta_j\hat{a}_j^\dag\hat{a}_j - \frac{1}{4}\Delta\chi \hat{\sigma}_z + \\
    \sum_{n=0}^1 \Theta_n' \ket{n}\bra{n} (\hat{a}^\dag_1 + \hat{a}_1) + \Omega' \hat{\sigma}_x (\hat{a}^\dag_2 + \hat{a}_2), \label{eq:2dion}
\end{multline}
where $\Delta\chi$ is the energy difference between qubit states induced by a time-independent AC Stark shift,
$\Theta_n'$ is proportional to the time-dependent AC Stark shift of state $n$ and $\Omega'$ is proportional to the Rabi frequency of the internal states of the ion (details in Appendix~A).

The initialization, evolution, and measurement steps of a pyrazine simulation are illustrated in Fig.~\ref{fig:scheme}a--c.
This simulation can be simplified by working in a displaced frame, which can transform the tuning term (third term in Eq.~\ref{eq:2dion}) to either
$\sigma_z (\hat{a}_1^\dag + \hat{a}_1)$ or $\ket{1}\bra{1} (\hat{a}_1^\dag + \hat{a}_1)$, which are more easily implemented in trapped ions (details in Appendix~B). To compensate for the frame displacement, Fig.~\ref{fig:scheme}a shows an initial wavefunction displacement.

\section{Analog Trotterization}
When simulating larger systems, it may be impractical to implement all VC interactions simultaneously on an MQB simulator, because doing so would require a separate light-matter interaction for each term in Eq.~\ref{eq:vcham}.
Instead, the evolution of the wavefunction can be discretized with a Suzuki-Trotter expansion~\cite{lloyd95} by applying sets of terms in sequence for short
times (Fig.~\ref{fig:scheme}d), allowing the same hardware to implement multiple interactions. A consequence of Trotterization is a trade-off
between the simulation accuracy and the size of Trotter steps; for example, in the first-order
Suzuki-Trotter expansion, the error scales with the square of the Trotter time step, $\Delta t^2$. Furthermore, as in many analog simulations, the problem is complicated by the fact
that the Hamiltonian of an MQB simulator is non-zero in the absence of interactions.
For example, the vibrational degrees of freedom in trapped ions continue to oscillate whether or not they are perturbed by a laser field.

Here, we describe two approaches to analog Trotterization. We assume that the base Hamiltonian $\hat{H}_0$ cannot be turned off, while interaction Hamiltonians $\hat{H}_k$, for $1\le k\le M$, can be turned on and off on demand.

The rescaling approach relies on the fact that, in an analog simulation, the total Hamiltonian can be scaled by a constant factor, changing only the speed of the simulation. The individual terms can also be rescaled, as long as the total Hamiltonian remains proportional to the molecular Hamiltonian. If we re-express the full Hamiltonian as
$
    \hat{H} = \hat{H}_0 + \sum_{k=1}^M \hat{H}_k = \sum_{k=1}^M (\hat{H}_0 / M + \hat{H}_k), \label{eq:trot}
$
the first-order evolution (for timestep $\Delta t$) is
$
    \hat{U}_1(\Delta t) = \prod_{k=1}^M e^{-i(\hat{H}_0 / M + \hat{H}_k)\Delta t}.
$
Essentially, the evolution by $\hat{H}_0$ is slowed down relative to the interaction terms to compensate for the continuous evolution under $\hat{H}_0$ while the $\hat{H}_k$ are implemented in turn.
Higher-order Suzuki-Trotter expansions can be used to reduce the Trotterization error.

The rewinding approach treats $\hat{H}_0$ and the interaction terms on the same timescale (i.e., without rescaling), instead reversing the excessive evolution under $\hat{H}_0$. This approach assumes that time evolution under $-\hat{H}_0$ can be implemented between Trotter steps. At first order, this scheme is given by
$
    \hat{U}_1(\Delta t) = e^{-i \hat{H}_0 \Delta t} \prod_{k=1}^M\big( e^{i \hat{H}_0 \Delta t} e^{-i(\hat{H}_0 + \hat{H}_k)\Delta t} \big),
$
i.e., by correcting the evolution by $\exp(i\hat{H}_0 \Delta t)$ after each Trotter step, which can be thought of as implementing another unitary of the form $\exp(-i(-2\hat{H}_0) \Delta t)$ while the base Hamiltonian is always on.

\begin{figure}
    \centering
    \includegraphics[width=\columnwidth]{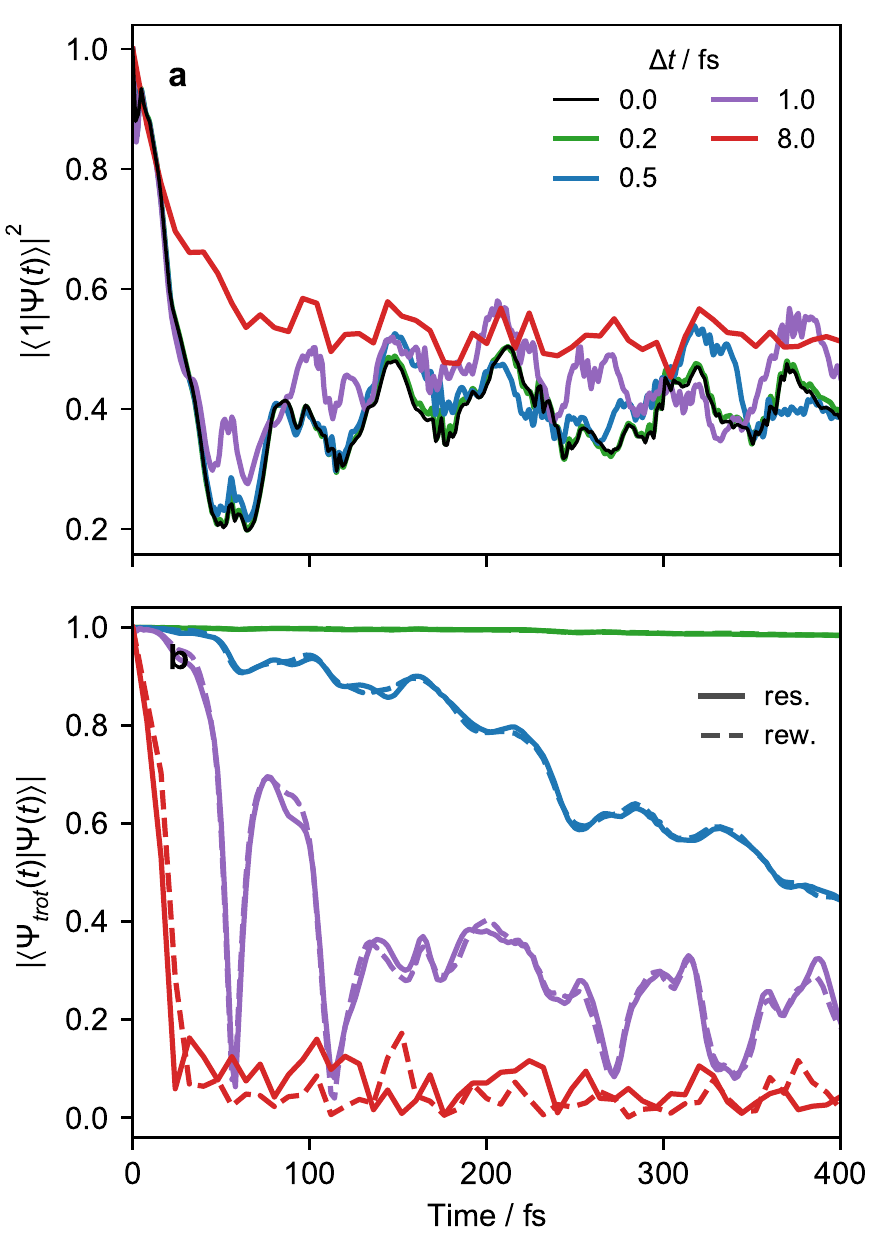}
    \caption{Evaluation of Trotterization approaches using the pyrazine LVC Hamiltonian. (\textbf{a}) Electronic populations as a function of time with different Trotter steps $\Delta t$, using the rescaling approach. (\textbf{b}) The rescaling (`res.') and rewinding (`rew.') approaches result in comparable fidelities (overlaps of the Trotterized wavefunction, $\ket{\Psi_\mathrm{trot}(t)}$, with the exact wavefunction, $\ket{\Psi(t)}$, with $\Delta t = 0$).}
    \label{fig:fid}
\end{figure}

The two approaches have similar performance, as can be seen in the example of a two-mode system, where
\begin{align}
    \hat{U}_1^\mathrm{res.}(\Delta t) &= e^{-i(\hat{H}_0 / 2 + \hat{H}_2)\Delta t} e^{-i(\hat{H}_0 / 2 + \hat{H}_1)\Delta t}, \\
    \hat{U}_1^\mathrm{rew.}(\Delta t) &= e^{-i(\hat{H}_0 + \hat{H}_2)\Delta t} e^{i\hat{H}_0\Delta t} e^{-i(\hat{H}_0 + \hat{H}_1)\Delta t},
\end{align}
where `$\mathrm{res.}$' and `$\mathrm{rew.}$' denote rescaling and rewinding, respectively. For the pyrazine example, we let $\hat{H}_0 = \sum_{j=1}^2 \delta_j \hat{a}_j^\dag \hat{a}_j$, $\hat{H}_1 = \sum_{n=0}^1 \Theta_n' \ket{n}\bra{n}(\hat{a}_1^\dag + \hat{a}_1) - \Delta\chi_1\hat{\sigma}_z/4$ and $\hat{H}_2 = \Omega'\hat{\sigma}_x(\hat{a}_2^\dag + \hat{a}_2) - \Delta\chi_2\hat{\sigma}_z/4$, with $\Delta\chi_1 = \Delta\chi_2 = \Delta\chi/2$. The effect of Trotterization is shown in Fig.~\ref{fig:fid}: as expected, increasing the Trotter step $\Delta t$ increases the error in the electronic population (Fig.~\ref{fig:fid}a). The fidelities (Fig.~\ref{fig:fid}b) show larger errors;
however, fidelities of both rescaling and rewinding approaches are nearly identical, indicating that either may be used for simulations.

A difference between the two approaches arises in the limit of many interaction terms, where one approach may be preferred over another depending on experimental limitations; the rescaling approach requires fewer operations, while the rewinding approach ensures that all components evolve at the same rate.

\section{Open quantum systems}
MQB devices also allow the simulation of open quantum systems---such as molecules in the condensed phase---with minimal overhead, using existing dissipation processes in the simulator. In general, open-system simulation
is even more computationally expensive on classical computers, as it involves the evolution of a density matrix rather than a state vector.
Classical MCTDH simulations have been achieved with weak coupling to baths, but increasing the number of bath modes quickly becomes intractable~\cite{worth08,wang15}.
By contrast, dissipation on an MQB simulator could be implemented using available environmental couplings. For example,
analog trapped-ion devices have simulated electronic dissipation with optical pumping~\cite{barreiro11} and
vibrational dissipation with sympathetic laser cooling~\cite{lemmer18,schlawin20}.

Here, we show how ion-trap MQB simulators could simulate
system-bath coupling with a desired coupling strength and effective temperature (Fig.~\ref{fig:scheme}e).
While simulating arbitrary system-bath interactions would be complicated, we focus on the leading contribution to dissipation in molecules in a thermal bath, coupling to vibrational modes~\cite{Nitzan2006}.
Our approach extends system-bath simulations using laser cooling~\cite{schlawin20},
and requires a maximum number of interactions that scales linearly with the number of modes coupled to the environment.

The conditions for a molecular simulation at a finite temperature are commonly given by the coupling of a constant-temperature bath of harmonic oscillators to the molecular modes to represent inter-molecular collisions~\cite{Nitzan2006,wang15}.
We consider a bath at temperature $T$ with jump operators $(\gamma_j(\bar{n}_j + 1))^{1/2}\hat{a}_j$ for cooling and $(\gamma_j\bar{n}_j)^{1/2}\hat{a}_j^\dag$ for heating of mode $j$, where $\gamma_j$ are the system-bath coupling strengths and $\bar{n}_j = (\exp(\omega_j/k_B T) - 1)^{-1}$ are the Bose-Einstein occupation numbers of the bath. This gives the master equation
\begin{align}
    \frac{\partial\hat{\rho}}{\partial t} = -i[\hat{H}, \hat{\rho}] + \sum_j \gamma_j\big((\bar{n}_j + 1)\mathcal{D}[\hat{a}_j] + \bar{n}_j\mathcal{D}[\hat{a}_j^\dag] \big)\hat{\rho},\label{eq:cooling}
\end{align}
where $\hat{\rho}$ is the density operator of the system and $\mathcal{D}[\hat{L}_j]$ is the Linblad superoperator for the jump operator $\hat{L}_j$, given by $\mathcal{D}[\hat{L}_j]\hat{\rho} = \hat{L}_j\hat{\rho}\hat{L}_j^\dag - \frac{1}{2}\{\hat{L}_j^\dag\hat{L}_j, \hat{\rho}\}$.

\begin{figure}[t]
    \centering
    \includegraphics[width=\columnwidth]{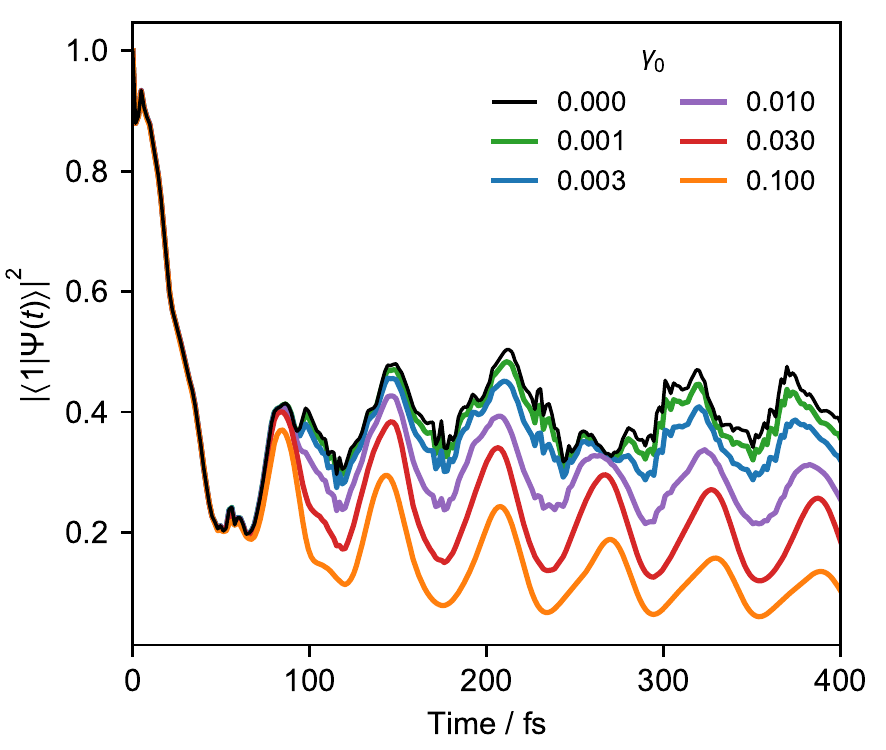}
    \caption{Electronic populations of pyrazine, using the LVC Hamiltonian and different system-bath couplings $\gamma_0$, all at a temperature of 300 K. As $\gamma_0$ is increased, the details of the closed-system result ($\gamma_0 = 0$) are lost and the coherent oscillations are damped.}
    \label{fig:oqs}
\end{figure}

Similar time evolution can be obtained using laser cooling, which is used in
ion-trap quantum computation to cool the ions to their ground motional state. Applied lasers both cool and heat; in the resolved sideband limit, the combined evolution due to laser cooling and heating takes the form \cite{stenholm86,lemmer18}
\begin{align}
    \frac{\partial\hat{\rho}}{\partial t} &= -i[\hat{H}, \hat{\rho}] + \sum_j \big( A_j^- \mathcal{D}[\hat{a}_j] + A_j^+ \mathcal{D}[\hat{a}_j^\dag] \big)\hat{\rho},\label{eq:heatcool}
\end{align}
where
\begin{align}
    A_j^{\pm} &= \eta_j^2 \Gamma_j \left( B_j(\Delta \pm \omega_j^\mathrm{ion}) + \alpha B_j(\Delta)\right), \\
    B_j(\Delta) &= \frac{\Omega_0^2}{\Gamma_j^2 + 4\Delta^2},
\end{align}
where $\eta_j$ is the Lamb-Dicke parameter, $\Gamma_j$ is the spontaneous emission rate of the carrier transition (i.e., the $\ket{0} \rightarrow \ket{1}$ transition),
$\Delta$ is the detuning from the carrier transition, $\alpha$
is the angular factor for spontaneous emission ($\alpha = 2/5$ for dipole transitions~\cite{stenholm86}), and $\Omega_0 \ll \Gamma_j$ is the Rabi frequency of the carrier transition~\cite{stenholm86}.

Comparing Eqs.~\ref{eq:cooling} and~\ref{eq:heatcool} shows that laser cooling and heating can simulate molecular thermal dissipation if the experimental parameters are chosen so that $\gamma_j = A_j^- - A_j^+$ and $\bar{n}_j = A_j^+ / (A_j^- - A_j^+)$, which can be achieved in two steps. First, because $A_j^+ / (A_j^- - A_j^+)$ depends only on $\Delta$ and $\Gamma_j$, those two parameters can be chosen to simulate a specific $\bar{n}_j$ and, therefore, the temperature. Second, $A_j^- - A_j^+$ is proportional to the square of the remaining free parameter, $\Omega_0$, which depends on the tunable laser intensity and can therefore be used to set the system-bath interaction $\gamma_j$ for each mode $j$.

The experimental couplings $\gamma_j$ must be scaled to simulator frequencies, as was done for $\hat{H}$. Such a rescaling changes the timescale in Eq.~\ref{eq:cooling}, giving Rabi frequencies in Hz--kHz instead of GHz--THz. By contrast, the occupation numbers $\bar{n}_j$ are dimensionless and require no scaling.
The difference between molecular and trapped-ion frequencies results in a difference in their vibrational temperatures because the invariable $\bar{n}_j$ depends only on $\omega_j/k_B T$. For example, a simulation at $\SI{300}{K}$ corresponds to an ion-trap vibrational temperature of $\sim\SI{40}{\micro K}$, which is readily achievable in ion traps~\cite{Diedrich.1989}.

Fig.~\ref{fig:oqs} shows an example of weak system-bath coupling with the LVC model of pyrazine at \SI{300}{K}. The simulation assumes the Ohmic system-bath couplings that are widely used in chemical dynamics~\cite{kuhl02}, $\gamma_j = \gamma_0 \omega_j e^{-\omega_j/\omega_0}$, where $\gamma_0$ is constant and $\omega_0$ is the high-frequency cutoff.

Our scheme requires at most one laser-cooling interaction per mode. However, because chemical dynamics often takes place at low temperatures ($k_B T \ll \omega_j$), it may be possible to simultaneously cool all the vibrational modes with only a single broadband cooling laser. Even at room temperature, $\bar{n}_j \ll 1$ for typical molecular vibrations with frequencies of tens of THz. In that limit, only the cooling terms in Eq.~\ref{eq:cooling} are significant,
$\partial\hat{\rho}/\partial t = -i[\hat{H}, \hat{\rho}] + \sum_j \gamma_j\mathcal{D}[\hat{a}_j]\hat{\rho}.$
If we also assume that all the $\gamma_j$ are comparable (which is true for a slowly varying Ohmic spectral density and comparable values of $\omega_j$), we arrive at a master equation that can be simulated with just one broadband cooling laser, which can cool all the modes simultaneously~\cite{Lechner.2016}. For example, repeating the simulation of Fig.~\ref{fig:oqs} with a single cooling laser and equal $\gamma_j$ leads to imperceptible changes in the populations shown.

\section{Limitations}
The main limitation of our approach---and of any analog quantum simulation---is the accumulation of simulation errors due to decoherence, dissipation, and the absence of error correction. As a result, hardware limitations place limits on the size of molecular systems that can be simulated and the duration of those simulations. Of course, the same was true of analog classical simulations, which were indispensable tools in the early decades of classical computation. Even with these limitations, early MQBs are likely to be useful, especially if the decoherence is harnessed as a resource.

The principal sources of error for MQB simulation will be decoherence and thermalization, but the subspace most affected may depend on the architecture.
For example, while trapped-ion simulators have electronic coherence times of seconds or longer, motional decoherence (typically dephasing) will limit simulations because it sets in within 1--100~ms \cite{lucas07,talukdar16}.
By contrast, in cQED, the limiting factor is likely to be the qudit coherence time, typically 50--100 \si{\micro s} and shorter than the coherence time of the cavity modes, which is limited by cavity losses and may be 2 ms or more~\cite{wang09,blais20}.

Whether an MQB simulator will be useful depends on the ratio of its decoherence time
and the necessary simulation time at realistic experimental frequencies. A simulation requires a scaling factor $F$ that relates molecular and simulator frequencies,
$\omega_\mathrm{sim} = F\omega_\mathrm{mol}$.
In order to complete a simulation for a time $t_\mathrm{max}$ (in molecular timescales)
within a decoherence time $\tau_d$, $F$ must exceed $t_\mathrm{max} / \tau_d$.
If, in addition, the time evolution is broken into $M$
Trotter components, the lower bound of $F$ is
$F_\mathrm{min} = M t_\mathrm{max} / \tau_d$.
There are also two upper limits on $F$ determined by experimental conditions: first, as the ratio of the maximum experimentally achievable frequency or coupling to the corresponding molecular parameter, e.g., $F^{(1)}_\mathrm{max} = \Theta' / \lambda$ for a coupling interaction;
and second, in a Trotterized simulation, as the ratio of the
maximum acceptable molecular-scale timestep to the minimum timestep achievable
experimentally, $F^{(2)}_\mathrm{max} = \Delta t_\mathrm{mol} / \Delta t_\mathrm{sim}$.
The simulation will be completed within the decoherence time if $F$ can be chosen to meet all these constraints, i.e., if both  $F^{(1)}_\mathrm{max}$ and $F^{(2)}_\mathrm{max}$ are larger than $F_\mathrm{min}$.
To minimize decoherence, $F$ should ideally be set to the upper bound.

The constraints above can be relaxed if the native dissipation in an MQB simulator can be harnessed to simulate dynamics in open quantum systems, which are often of greater practical relevance than isolated molecules. For example, decoherence and dissipation of
bosonic modes might be tuned (e.g., via temperature) to values that simulate elastic and inelastic collisions in condensed phases or at high pressures. For such a simulation, care must be taken to consider the relative dissipation rates
of electronic and vibrational modes, particularly for cQED simulators where the relevant internal and bosonic frequencies have similar magnitudes.

The limits of a simulation also depend on the observable of interest, meaning that many simulations will be able to run longer than worst-case estimates from fidelities.
In general, errors in analog simulation depend strongly on the observable, with global properties being more robust~\cite{sarovar17,poggi20}. For example, in Fig.~\ref{fig:fid} with a Trotter step of 0.5~fs (blue lines), the population deviates by at most 0.05 in 300~fs, whereas the fidelity rapidly decays to less than 0.6 in the same time. Fortunately, most chemically interesting molecular observables are global properties, rather than properties of individual vibronic eigenstates, which would allow for greater error tolerance than fidelities would suggest.

Finally, in very large systems, the number of bosonic modes could be limited by the experimental apparatus.
In trapped ions, the frequencies of the modes become increasingly spectrally congested
as the number of ions increases, which may eventually  lead to cross-talk due to multiple modes interacting with a single laser freqency~\cite{lee05,friis18}. So far, ion trap experiments have been performed with as many as 40 resolved modes~\cite{friis18}. In cQED, resonators are physically coupled to qudits, placing spatial
limitations in connecting many resonators a single qudit. However, if multimode resonators~\cite{sundaresan15,kollar19} are used, their total number may be suitable for medium-scale simulations. Therefore, the current technological capabilities are more than
enough for near-term demonstration and will undoubtedly improve as technology develops.

\section{Conclusions}
The one-to-one correspondence of molecular and simulator degrees of freedom gives MQB simulators a clear advantage in the simulation of non-adiabatic chemical dynamics over both classical and qubit-based schemes.
For an increasing number of vibrational modes, our approach requires linearly scaling quantum hardware and polynomially many interaction terms, in either isolated or open molecular systems.
For large systems, the number of simultaneous interactions can be significantly reduced with a Suzuki-Trotter expansion, allowing the same driving fields to be reused.
Scaling the frequencies of the VC Hamiltonian to the MQB simulator also offers a significant advantage in time resolution relative to ultrafast spectroscopy experiments.
Many of the tools necessary for MQB simulation have been developed in the context of digital quantum computation and can be repurposed for analog simulation, making MQB simulators a promising platform for quantum advantage in the simulation of chemical dynamics.
Indeed, we expect that dynamical simulations will more readily achieve quantum advantage than solving time-independent electronic-structure problems.

\section*{Numerical methods}
Simulations of the closed-system LVC model of pyrazine~\cite{kuhl02} were performed using NumPy~\cite{numpy} and SciPy~\cite{scipy} with the number
of vibrational eigenstates truncated to 20 for both modes ($\nu_{6a}$ and $\nu_{10a}$). Open-system dynamics used the Linblad master-equation solver in QuTiP~\cite{qutip}.

\section*{Appendix A: Light-matter interactions in trapped-ion simulators}
\label{app:light}
Here, we follow ref.~\citenum{lee05} to derive the light-matter interactions required to map an LVC Hamiltonian onto an ion-trap Hamiltonian.

Tuning terms in the LVC Hamiltonian can be generated using
two non-copropagating lasers, which can induce Raman transitions that couple the internal states of an $N$-level trapped ion with its vibrational mode $j$ (frequency $\omega_j^\mathrm{ion}$). If the frequency between the two lasers is $\omega_j^\mathrm{ion} - \delta_j$, the resulting interaction-picture Hamiltonian becomes
\begin{align}
    \hat{H}_I^t &= \frac{1}{2}\eta_j D_j' \sum_{n} \Theta_{n,j} \ket{n}\bra{n} (\hat{a}^\dag_j e^{i(\delta_j t - \phi)} + \hc), \label{eq:tun}
\end{align}
where $\eta_j$ is the Lamb-Dicke parameter, $D_j'$ is the Debye-Waller factor, and $\Theta_{n,j}$ is the time-dependent AC Stark shift of state $n$.
The interaction of Eq.~\ref{eq:tun} is well established; in particular, the two-qubit generalisation is widely used to implement the phase gate $\hat{\sigma}_z\otimes \hat{\sigma}_z$~\cite{leibfried03}.
This form of the Hamiltonian assumes that the intermediate state $\ket{e}$ can be adiabatically eliminated, i.e., that the global detuning $\Delta$ of $\omega_\alpha$ and $\omega_\beta$
from $\ket{e}$ is much larger than the linewidth $\gamma_e$ of $\ket{e}$, and that the detuning $\delta_j$ is small relative to the ion trap frequency $\omega_j^\mathrm{ion}$. The laser field also causes a time-independent AC Stark
shift for each state, $\chi^{(n)}$, which modulates the internal state energies and can be experimentally tuned by $\Delta$~\cite{leibfried03}. Without loss of
generality, the phase $\phi$ can be dropped as it corresponds to an arbitrary initial phase-space angle.

Coupling terms in the LVC Hamiltonian can be generated if, instead, the difference between the laser frequencies is $\Delta E_{n,m}^\mathrm{ion} \pm (\omega_k^\mathrm{ion} - \delta_k)$, where $\Delta E_{n,m}^\mathrm{ion}$ is the energy difference between internal states $n$ and $m$, and the plus or minus determining whether the interaction corresponds to a blue- or red-sideband transition. The sum of red and blue sideband terms gives a Hamiltonian of the form
\begin{multline}
    \hat{H}_I^c = \frac{1}{2}\eta_k D_k' \Omega_{n,m,k} (\ket{n}\bra{m}e^{i\phi_S} + \hc) \\
    \times (\hat{a}^\dag_k e^{i(\delta_k t - \phi_M)} + \hc), \label{eq:coup}
\end{multline}
where $\Omega_{n,m,k}$ is the base Rabi frequency of the internal states, $\phi_S$ is the spin-phase of the ion resulting from the sum of the two transitions, and
$\phi_M$ is the phase difference between the red and blue sidebands. As with the interaction above, this interaction is widely used, especially to generate the M{\o}lmer-S{\o}rensen entangling gate~\cite{molmer99,sorensen99}. Setting $\phi_S = 0$ results in an interstate coupling in the same form as in the LVC Hamiltonian. As with $\phi$ before, the phase $\phi_M$ can be dropped.

The total interaction Hamiltonian is obtained by adding an instance of Eq.~\ref{eq:tun} for each tuning mode and an instance of Eq.~\ref{eq:coup} for each coupling mode. Once the time-independent AC Stark shifts $\chi_n$ are also included, the total Hamiltonian becomes
\begin{align}
    \hat{H}_I^\mathrm{MQB} &= \sum_{j \in \mathbf{t}} \sum_n \Theta_{n,j}' \ket{n}\bra{n} (\hat{a}^\dag_j e^{i\delta_j t} + \hc) \nonumber\\
    &\quad+ \sum_{k \in \mathbf{c}} \sum_{n \neq m} \Omega_{n,m,k}' \ket{n}\bra{m} (\hat{a}^\dag_k e^{i\delta_k t} + \hc)\nonumber\\
    &\quad + \frac{1}{2}\sum_n \chi_n \ket{n}\bra{n},
\end{align}
where $\Theta_{n,j}' = \eta_j D_j' \Theta_{n,j}/2$, $\Omega_{n,m,k}' = \eta_k D_k' \Omega_{n,m,k} / 2$ and the sets $\mathbf{t}$
and $\mathbf{c}$ are indices of tuning and coupling modes, respectively. This Hamiltonian
can be transformed into a rotating frame to give
\begin{align}
    \hat{H}^\mathrm{MQB} &= \sum_j \delta_j \hat{a}_j^\dag \hat{a}_j + \frac{1}{2}\sum_n \chi_n\ket{n}\bra{n} \nonumber\\
    &\quad+ \sum_{j \in \mathbf{t}} \sum_n \Theta_{n,j}' \ket{n}\bra{n} (\hat{a}^\dag_j + \hat{a}_j) \nonumber\\
    &\quad+ \sum_{k \in \mathbf{c}}\sum_{n \neq m} \Omega_{n,m,k}' \ket{n}\bra{m}(\hat{a}^\dag_k + \hat{a}_k).
    \label{eq:full}
\end{align}

The results above are general for any number of internal states and any number of vibrational modes. Although a single ion has only three vibrational modes, additional modes can be obtained by adding ancillary, optically inactive ions into the trap. The ancillary ions can be made optically inactive in multiple ways, including the use of different isotopes~\cite{Hempel.2013} or by pumping them into long-lived shelved states~\cite{Hosaka.2009, Edmunds.2020b}.

For a two-state system ($n,m \in \{0, 1\}$ with two modes ($\mathbf{t} = \{1\}$, $\mathbf{c} = \{2\}$), Eq.~\ref{eq:full} simplifies to Eq.~\ref{eq:2dion} of the main text.

\section*{Appendix B: Displaced vibronic coupling Hamiltonians}
\label{app:disp}
For experimental implementation of VC Hamiltonian simulations, it may be convenient to transform the elements of the Hamiltonian into a different form. In particular, the interstate coupling factors of the form $\sum_{n\le m}c_j^{(n,m)}(\ket{n}\bra{m}+\hc)$ can be replaced with more easily implemented operators, such as $\hat{\sigma}_x$ or $\hat{\sigma}_z$.

One way to do this is by a coherent displacement of the Hamiltonian along a single mode. This corresponds to the action of a displacement operator,
$\hat{D}_k(\beta) = \exp(-i\beta\hat{P}_k)$ for mode $k$ and real $\beta$, where
\begin{align}
\hat{D}_k(\beta) \hat{Q}_k \hat{D}_k^\dag(\beta) = \hat{Q}_k - \beta
\end{align}
is a change in position by $\beta$ along mode $k$.
For a general VC Hamiltonian (Eq.~\ref{eq:vcham}), the displacement gives
\begin{align}
\hat{H}'_\mathrm{VC} &= \hat{D}_k \hat{H}_\mathrm{VC} \hat{D}_k^\dag = \hat{H}_\mathrm{VC} - \omega_k\beta\hat{Q}_k + \frac{1}{2}\omega_k\beta^2 \nonumber\\
&\quad- \sum_{n,m} \left(\beta c_k^{(n,m)} - \beta^2 c_{k,k}^{(n,m)} \phantom{\sum_j}\right.\nonumber\\
&\quad\left.+ 2\beta\sum_j c_{k,j}^{(n,m)} \hat{Q}_j + \cdots\right)\ket{n}\bra{m}.
\end{align}
Dropping the constant term $\omega_k\beta^2 / 2$ and truncating to linear terms ($c_{k,j,...}^{(n,m)} = 0$) yields the displaced LVC Hamiltonian,
\begin{align}
\hat{H}'_\mathrm{LVC} = \hat{H}_\mathrm{LVC} - \sum_{n,m} \left(\beta c_k^{(n,m)} + \beta\omega_k\delta_{n,m}\hat{Q}_k\right)\ket{n}\bra{m},
\end{align}
where $\delta_{n,m}$ is a Kronecker delta, meaning only the diagonal elements ($n = m$) of first-order terms are affected by displacement.

In the two-state case ($n,m \in \{0, 1\}$), the electronic components of the LVC Hamiltonian can be expressed in terms of Pauli operators to yield
\begin{multline}
\hat{H}_\mathrm{LVC} = \frac{1}{2}\sum_j \omega_j \left(\hat{Q}_j^2 + \hat{P}_j^2\right) - \frac{1}{2} \Delta E\hat{\sigma}_z + W_0\hat{\sigma}_x \\
+ \sum_j \left(\bar{\kappa}_j - \frac{1}{2}\Delta\kappa_j\hat{\sigma}_z + \lambda_j\hat{\sigma}_x\right) \hat{Q}_j, \label{eq:2stlvc}
\end{multline}
where $\Delta E = c_0^{(1,1)} - c_0^{(0,0)}$, $W_0 = c_0^{(0,1)}$, $\bar{\kappa}_j = (c_j^{(1,1)} + c_j^{(0,0)}) / 2$, $\Delta\kappa_j = c_j^{(1,1)} - c_j^{(0,0)}$ and $\lambda_j = c_j^{(0,1)}$. After displacement along mode $k$,
\begin{align}
\hat{H}'_\mathrm{LVC} = \hat{H}_\mathrm{LVC} - \beta \left(- \frac{1}{2}\Delta\kappa_k\hat{\sigma}_z + \lambda_k\hat{\sigma}_x + \omega_k\hat{Q}_k\right),
\end{align}
where the constant energy $\beta\bar{\kappa}_k$ is excluded as it only contributes a global phase. This is equivalent to making the replacements
$\Delta E \rightarrow \Delta E - \beta\Delta\kappa_k$, $W_0 \rightarrow W_0 - \beta\lambda_k$ and $\bar{\kappa}_k \rightarrow \bar{\kappa}_k - \beta\omega_k$,
which could be used to simplify the Hamiltonian. For example, choosing $\beta = \bar{\kappa}_k / \omega_k$ would remove the $\mathbb{1} \otimes \hat{Q}_k$ term from Eq.~\ref{eq:2stlvc}.

The dynamics simulated with the displaced VC Hamiltonian is
exactly equivalent (i.e., produces the same observables) to that of the original model as long as the initial wavefunction is displaced by an equal amount, which is achievable for common bosonic simulators~\cite{Hempel.2013,Vlastakis.2013}.

\section*{Acknowledgements}
We acknowledge valuable discussion with Alistair Milne, Ramil Nigmatullin, Ting Rei Tan, and Joel Yuen-Zhou.
We were supported by a Westpac Scholars Trust Research Fellowship, by the Lockheed Martin Corporation, by the Australian Government's Defence Science and Technology Group, by the United States Office of Naval Research Global (N62909-20-1-2047), and by the University of Sydney Nano Institute Grand Challenge \textit{Computational Materials Discovery}.
We were supported by computational resources and assistance from the National Computational Infrastructure (NCI) and the University of Sydney’s computing cluster Artemis.


\begin{thebibliography}{78}%
\makeatletter
\providecommand \@ifxundefined [1]{%
 \@ifx{#1\undefined}
}%
\providecommand \@ifnum [1]{%
 \ifnum #1\expandafter \@firstoftwo
 \else \expandafter \@secondoftwo
 \fi
}%
\providecommand \@ifx [1]{%
 \ifx #1\expandafter \@firstoftwo
 \else \expandafter \@secondoftwo
 \fi
}%
\providecommand \natexlab [1]{#1}%
\providecommand \enquote  [1]{``#1''}%
\providecommand \bibnamefont  [1]{#1}%
\providecommand \bibfnamefont [1]{#1}%
\providecommand \citenamefont [1]{#1}%
\providecommand \href@noop [0]{\@secondoftwo}%
\providecommand \href [0]{\begingroup \@sanitize@url \@href}%
\providecommand \@href[1]{\@@startlink{#1}\@@href}%
\providecommand \@@href[1]{\endgroup#1\@@endlink}%
\providecommand \@sanitize@url [0]{\catcode `\\12\catcode `\$12\catcode
  `\&12\catcode `\#12\catcode `\^12\catcode `\_12\catcode `\%12\relax}%
\providecommand \@@startlink[1]{}%
\providecommand \@@endlink[0]{}%
\providecommand \url  [0]{\begingroup\@sanitize@url \@url }%
\providecommand \@url [1]{\endgroup\@href {#1}{\urlprefix }}%
\providecommand \urlprefix  [0]{URL }%
\providecommand \Eprint [0]{\href }%
\providecommand \doibase [0]{http://dx.doi.org/}%
\providecommand \selectlanguage [0]{\@gobble}%
\providecommand \bibinfo  [0]{\@secondoftwo}%
\providecommand \bibfield  [0]{\@secondoftwo}%
\providecommand \translation [1]{[#1]}%
\providecommand \BibitemOpen [0]{}%
\providecommand \bibitemStop [0]{}%
\providecommand \bibitemNoStop [0]{.\EOS\space}%
\providecommand \EOS [0]{\spacefactor3000\relax}%
\providecommand \BibitemShut  [1]{\csname bibitem#1\endcsname}%
\let\auto@bib@innerbib\@empty
%</preamble>
\bibitem [{\citenamefont {Domcke}\ \emph {et~al.}(2004)\citenamefont {Domcke},
  \citenamefont {Yarkony},\ and\ \citenamefont {K{\" o}ppel}}]{domcke04}%
  \BibitemOpen
  \bibinfo {editor} {\bibfnamefont {W.}~\bibnamefont {Domcke}}, \bibinfo
  {editor} {\bibfnamefont {D.~R.}\ \bibnamefont {Yarkony}}, \ and\ \bibinfo
  {editor} {\bibfnamefont {H.}~\bibnamefont {K{\" o}ppel}},\ eds.,\ \href@noop
  {} {\emph {\bibinfo {title} {{Conical Intersections: Electronic Structure,
  Dynamics \& Spectroscopy}}}},\ \bibinfo {series} {Adv.\ Ser.\ Phys.\ Chem.},
  Vol.~\bibinfo {volume} {15}\ (\bibinfo  {publisher} {World Scientific},\
  \bibinfo {address} {Singapore},\ \bibinfo {year} {2004})\BibitemShut
  {NoStop}%
\bibitem [{\citenamefont {Cederbaum}\ \emph {et~al.}(1981)\citenamefont
  {Cederbaum}, \citenamefont {K{\" o}ppel},\ and\ \citenamefont
  {Domcke}}]{cederbaum81}%
  \BibitemOpen
  \bibfield  {author} {\bibinfo {author} {\bibfnamefont {L.~S.}\ \bibnamefont
  {Cederbaum}}, \bibinfo {author} {\bibfnamefont {H.}~\bibnamefont {K{\"
  o}ppel}}, \ and\ \bibinfo {author} {\bibfnamefont {W.}~\bibnamefont
  {Domcke}},\ }\href {\doibase 10.1002/qua.560200828} {\bibfield  {journal}
  {\bibinfo  {journal} {Int. J. Quantum Chem.}\ }\textbf {\bibinfo {volume}
  {20}},\ \bibinfo {pages} {251} (\bibinfo {year} {1981})}\BibitemShut
  {NoStop}%
\bibitem [{\citenamefont {Worth}\ \emph {et~al.}(2008)\citenamefont {Worth},
  \citenamefont {Meyer}, \citenamefont {K{\" o}ppel}, \citenamefont
  {Cederbaum},\ and\ \citenamefont {Burghardt}}]{worth08}%
  \BibitemOpen
  \bibfield  {author} {\bibinfo {author} {\bibfnamefont {G.~A.}\ \bibnamefont
  {Worth}}, \bibinfo {author} {\bibfnamefont {H.-D.}\ \bibnamefont {Meyer}},
  \bibinfo {author} {\bibfnamefont {H.}~\bibnamefont {K{\" o}ppel}}, \bibinfo
  {author} {\bibfnamefont {L.~S.}\ \bibnamefont {Cederbaum}}, \ and\ \bibinfo
  {author} {\bibfnamefont {I.}~\bibnamefont {Burghardt}},\ }\href {\doibase
  10.1080/01442350802137656} {\bibfield  {journal} {\bibinfo  {journal} {Int.\
  Rev.\ Phys.\ Chem.}\ }\textbf {\bibinfo {volume} {27}},\ \bibinfo {pages}
  {569} (\bibinfo {year} {2008})}\BibitemShut {NoStop}%
\bibitem [{\citenamefont {Wang}(2015)}]{wang15}%
  \BibitemOpen
  \bibfield  {author} {\bibinfo {author} {\bibfnamefont {H.}~\bibnamefont
  {Wang}},\ }\href {\doibase 10.1021/acs.jpca.5b03256} {\bibfield  {journal}
  {\bibinfo  {journal} {J.\ Phys.\ Chem.\ A}\ }\textbf {\bibinfo {volume}
  {119}},\ \bibinfo {pages} {7951} (\bibinfo {year} {2015})}\BibitemShut
  {NoStop}%
\bibitem [{\citenamefont {Meng}\ and\ \citenamefont {Meyer}(2013)}]{meng13}%
  \BibitemOpen
  \bibfield  {author} {\bibinfo {author} {\bibfnamefont {Q.}~\bibnamefont
  {Meng}}\ and\ \bibinfo {author} {\bibfnamefont {H.-D.}\ \bibnamefont
  {Meyer}},\ }\href {\doibase 10.1063/1.4772779} {\bibfield  {journal}
  {\bibinfo  {journal} {J.\ Chem.\ Phys.}\ }\textbf {\bibinfo {volume} {138}},\
  \bibinfo {pages} {014313} (\bibinfo {year} {2013})}\BibitemShut {NoStop}%
\bibitem [{\citenamefont {Xie}\ \emph {et~al.}(2015)\citenamefont {Xie},
  \citenamefont {Zheng},\ and\ \citenamefont {Lan}}]{xie15}%
  \BibitemOpen
  \bibfield  {author} {\bibinfo {author} {\bibfnamefont {Y.}~\bibnamefont
  {Xie}}, \bibinfo {author} {\bibfnamefont {J.}~\bibnamefont {Zheng}}, \ and\
  \bibinfo {author} {\bibfnamefont {Z.}~\bibnamefont {Lan}},\ }\href {\doibase
  10.1063/1.4909521} {\bibfield  {journal} {\bibinfo  {journal} {J.\ Chem.\
  Phys.}\ }\textbf {\bibinfo {volume} {142}},\ \bibinfo {pages} {084706}
  (\bibinfo {year} {2015})}\BibitemShut {NoStop}%
\bibitem [{\citenamefont {Schulze}\ \emph {et~al.}(2016)\citenamefont
  {Schulze}, \citenamefont {Shibl}, \citenamefont {Al-Marri},\ and\
  \citenamefont {K{\" u}hn}}]{schulze16}%
  \BibitemOpen
  \bibfield  {author} {\bibinfo {author} {\bibfnamefont {J.}~\bibnamefont
  {Schulze}}, \bibinfo {author} {\bibfnamefont {M.~F.}\ \bibnamefont {Shibl}},
  \bibinfo {author} {\bibfnamefont {M.~J.}\ \bibnamefont {Al-Marri}}, \ and\
  \bibinfo {author} {\bibfnamefont {O.}~\bibnamefont {K{\" u}hn}},\ }\href
  {\doibase 10.1063/1.4948563} {\bibfield  {journal} {\bibinfo  {journal} {J.\
  Chem.\ Phys.}\ }\textbf {\bibinfo {volume} {144}},\ \bibinfo {pages} {185101}
  (\bibinfo {year} {2016})}\BibitemShut {NoStop}%
\bibitem [{\citenamefont {Feynman}(1982)}]{Feynman.1982}%
  \BibitemOpen
  \bibfield  {author} {\bibinfo {author} {\bibfnamefont {R.}~\bibnamefont
  {Feynman}},\ }\href@noop {} {\bibfield  {journal} {\bibinfo  {journal} {Int.
  J. Theor. Phys.}\ }\textbf {\bibinfo {volume} {21}},\ \bibinfo {pages} {467 }
  (\bibinfo {year} {1982})}\BibitemShut {NoStop}%
\bibitem [{\citenamefont {Lloyd}(1995)}]{lloyd95}%
  \BibitemOpen
  \bibfield  {author} {\bibinfo {author} {\bibfnamefont {S.}~\bibnamefont
  {Lloyd}},\ }\href {\doibase 10.1103/physrevlett.75.346} {\bibfield  {journal}
  {\bibinfo  {journal} {Phys.\ Rev.\ Lett.}\ }\textbf {\bibinfo {volume}
  {75}},\ \bibinfo {pages} {346} (\bibinfo {year} {1995})}\BibitemShut
  {NoStop}%
\bibitem [{\citenamefont {Nielsen}\ and\ \citenamefont
  {Chuang}(2000)}]{nielsen10}%
  \BibitemOpen
  \bibfield  {author} {\bibinfo {author} {\bibfnamefont {M.~A.}\ \bibnamefont
  {Nielsen}}\ and\ \bibinfo {author} {\bibfnamefont {I.~L.}\ \bibnamefont
  {Chuang}},\ }\href@noop {} {\emph {\bibinfo {title} {{Quantum Computation and
  Quantum Information}}}}\ (\bibinfo  {publisher} {Cambridge University
  Press},\ \bibinfo {address} {Cambridge, UK},\ \bibinfo {year}
  {2000})\BibitemShut {NoStop}%
\bibitem [{\citenamefont {Buluta}\ and\ \citenamefont
  {Nori}(2009)}]{Buluta2009}%
  \BibitemOpen
  \bibfield  {author} {\bibinfo {author} {\bibfnamefont {I.}~\bibnamefont
  {Buluta}}\ and\ \bibinfo {author} {\bibfnamefont {F.}~\bibnamefont {Nori}},\
  }\href@noop {} {\bibfield  {journal} {\bibinfo  {journal} {Science}\ }\textbf
  {\bibinfo {volume} {326}},\ \bibinfo {pages} {108} (\bibinfo {year}
  {2009})}\BibitemShut {NoStop}%
\bibitem [{\citenamefont {Blatt}\ and\ \citenamefont {Roos}(2012)}]{blatt2012}%
  \BibitemOpen
  \bibfield  {author} {\bibinfo {author} {\bibfnamefont {R.}~\bibnamefont
  {Blatt}}\ and\ \bibinfo {author} {\bibfnamefont {C.~F.}\ \bibnamefont
  {Roos}},\ }\href {\doibase 10.1038/nphys2252} {\bibfield  {journal} {\bibinfo
   {journal} {Nat.\ Phys.}\ }\textbf {\bibinfo {volume} {8}},\ \bibinfo {pages}
  {277} (\bibinfo {year} {2012})}\BibitemShut {NoStop}%
\bibitem [{\citenamefont {Aspuru-Guzik}\ and\ \citenamefont
  {Walther}(2012)}]{aspuru2012}%
  \BibitemOpen
  \bibfield  {author} {\bibinfo {author} {\bibfnamefont {A.}~\bibnamefont
  {Aspuru-Guzik}}\ and\ \bibinfo {author} {\bibfnamefont {P.}~\bibnamefont
  {Walther}},\ }\href@noop {} {\bibfield  {journal} {\bibinfo  {journal} {Nat.\
  Phys.}\ }\textbf {\bibinfo {volume} {8}},\ \bibinfo {pages} {285} (\bibinfo
  {year} {2012})}\BibitemShut {NoStop}%
\bibitem [{\citenamefont {Bloch}\ \emph {et~al.}(2012)\citenamefont {Bloch},
  \citenamefont {Dalibard},\ and\ \citenamefont {Nascimbène}}]{bloch2012}%
  \BibitemOpen
  \bibfield  {author} {\bibinfo {author} {\bibfnamefont {I.}~\bibnamefont
  {Bloch}}, \bibinfo {author} {\bibfnamefont {J.}~\bibnamefont {Dalibard}}, \
  and\ \bibinfo {author} {\bibfnamefont {S.}~\bibnamefont {Nascimbène}},\
  }\href {\doibase 10.1038/nphys2259} {\bibfield  {journal} {\bibinfo
  {journal} {Nat.\ Phys.}\ }\textbf {\bibinfo {volume} {8}},\ \bibinfo {pages}
  {267} (\bibinfo {year} {2012})}\BibitemShut {NoStop}%
\bibitem [{\citenamefont {Georgescu}\ \emph {et~al.}(2014)\citenamefont
  {Georgescu}, \citenamefont {Ashhab},\ and\ \citenamefont
  {Nori}}]{RevModPhys.86.153}%
  \BibitemOpen
  \bibfield  {author} {\bibinfo {author} {\bibfnamefont {I.~M.}\ \bibnamefont
  {Georgescu}}, \bibinfo {author} {\bibfnamefont {S.}~\bibnamefont {Ashhab}}, \
  and\ \bibinfo {author} {\bibfnamefont {F.}~\bibnamefont {Nori}},\ }\href
  {\doibase 10.1103/RevModPhys.86.153} {\bibfield  {journal} {\bibinfo
  {journal} {Rev. Mod. Phys.}\ }\textbf {\bibinfo {volume} {86}},\ \bibinfo
  {pages} {153} (\bibinfo {year} {2014})}\BibitemShut {NoStop}%
\bibitem [{\citenamefont {Schäfer}\ \emph {et~al.}(2020)\citenamefont
  {Schäfer}, \citenamefont {Fukuhara}, \citenamefont {Sugawa}, \citenamefont
  {Takasu},\ and\ \citenamefont {Takahashi}}]{schafer2020}%
  \BibitemOpen
  \bibfield  {author} {\bibinfo {author} {\bibfnamefont {F.}~\bibnamefont
  {Schäfer}}, \bibinfo {author} {\bibfnamefont {T.}~\bibnamefont {Fukuhara}},
  \bibinfo {author} {\bibfnamefont {S.}~\bibnamefont {Sugawa}}, \bibinfo
  {author} {\bibfnamefont {Y.}~\bibnamefont {Takasu}}, \ and\ \bibinfo {author}
  {\bibfnamefont {Y.}~\bibnamefont {Takahashi}},\ }\href {\doibase
  10.1038/s42254-020-0195-3} {\bibfield  {journal} {\bibinfo  {journal} {Nat.\
  Rev.\ Phys.}\ }\textbf {\bibinfo {volume} {2}},\ \bibinfo {pages} {411}
  (\bibinfo {year} {2020})}\BibitemShut {NoStop}%
\bibitem [{\citenamefont {Aspuru-Guzik}\ \emph {et~al.}(2005)\citenamefont
  {Aspuru-Guzik}, \citenamefont {Dutoi}, \citenamefont {Love},\ and\
  \citenamefont {Head-Gordon}}]{aspuruguzik05}%
  \BibitemOpen
  \bibfield  {author} {\bibinfo {author} {\bibfnamefont {A.}~\bibnamefont
  {Aspuru-Guzik}}, \bibinfo {author} {\bibfnamefont {A.~D.}\ \bibnamefont
  {Dutoi}}, \bibinfo {author} {\bibfnamefont {P.~J.}\ \bibnamefont {Love}}, \
  and\ \bibinfo {author} {\bibfnamefont {M.}~\bibnamefont {Head-Gordon}},\
  }\href {\doibase 10.1126/science.1113479} {\bibfield  {journal} {\bibinfo
  {journal} {Science}\ }\textbf {\bibinfo {volume} {309}},\ \bibinfo {pages}
  {1704} (\bibinfo {year} {2005})}\BibitemShut {NoStop}%
\bibitem [{\citenamefont {Lanyon}\ \emph {et~al.}(2010)\citenamefont {Lanyon},
  \citenamefont {Whitfield}, \citenamefont {Gillett}, \citenamefont {Goggin},
  \citenamefont {Almeida}, \citenamefont {Kassal}, \citenamefont {Biamonte},
  \citenamefont {Mohseni}, \citenamefont {Powell}, \citenamefont {Barbieri},
  \citenamefont {Aspuru-Guzik},\ and\ \citenamefont {White}}]{lanyon10}%
  \BibitemOpen
  \bibfield  {author} {\bibinfo {author} {\bibfnamefont {B.~P.}\ \bibnamefont
  {Lanyon}}, \bibinfo {author} {\bibfnamefont {J.~D.}\ \bibnamefont
  {Whitfield}}, \bibinfo {author} {\bibfnamefont {G.~G.}\ \bibnamefont
  {Gillett}}, \bibinfo {author} {\bibfnamefont {M.~E.}\ \bibnamefont {Goggin}},
  \bibinfo {author} {\bibfnamefont {M.~P.}\ \bibnamefont {Almeida}}, \bibinfo
  {author} {\bibfnamefont {I.}~\bibnamefont {Kassal}}, \bibinfo {author}
  {\bibfnamefont {J.~D.}\ \bibnamefont {Biamonte}}, \bibinfo {author}
  {\bibfnamefont {M.}~\bibnamefont {Mohseni}}, \bibinfo {author} {\bibfnamefont
  {B.~J.}\ \bibnamefont {Powell}}, \bibinfo {author} {\bibfnamefont
  {M.}~\bibnamefont {Barbieri}}, \bibinfo {author} {\bibfnamefont
  {A.}~\bibnamefont {Aspuru-Guzik}}, \ and\ \bibinfo {author} {\bibfnamefont
  {A.~G.}\ \bibnamefont {White}},\ }\href {\doibase 10.1038/nchem.483}
  {\bibfield  {journal} {\bibinfo  {journal} {Nat.\ Chem.}\ }\textbf {\bibinfo
  {volume} {2}},\ \bibinfo {pages} {106} (\bibinfo {year} {2010})}\BibitemShut
  {NoStop}%
\bibitem [{\citenamefont {Yung}\ \emph {et~al.}(2014)\citenamefont {Yung},
  \citenamefont {Casanova}, \citenamefont {Mezzacapo}, \citenamefont {McClean},
  \citenamefont {Lamata}, \citenamefont {Aspuru-Guzik},\ and\ \citenamefont
  {Solano}}]{yung14}%
  \BibitemOpen
  \bibfield  {author} {\bibinfo {author} {\bibfnamefont {M.-H.}\ \bibnamefont
  {Yung}}, \bibinfo {author} {\bibfnamefont {J.}~\bibnamefont {Casanova}},
  \bibinfo {author} {\bibfnamefont {A.}~\bibnamefont {Mezzacapo}}, \bibinfo
  {author} {\bibfnamefont {J.}~\bibnamefont {McClean}}, \bibinfo {author}
  {\bibfnamefont {L.}~\bibnamefont {Lamata}}, \bibinfo {author} {\bibfnamefont
  {A.}~\bibnamefont {Aspuru-Guzik}}, \ and\ \bibinfo {author} {\bibfnamefont
  {E.}~\bibnamefont {Solano}},\ }\href {\doibase 10.1038/srep03589} {\bibfield
  {journal} {\bibinfo  {journal} {Sci.\ Rep.}\ }\textbf {\bibinfo {volume}
  {4}},\ \bibinfo {pages} {3589} (\bibinfo {year} {2014})}\BibitemShut
  {NoStop}%
\bibitem [{\citenamefont {Peruzzo}\ \emph {et~al.}(2014)\citenamefont
  {Peruzzo}, \citenamefont {McClean}, \citenamefont {Shadbolt}, \citenamefont
  {Yung}, \citenamefont {Zhou}, \citenamefont {Love}, \citenamefont
  {Aspuru-Guzik},\ and\ \citenamefont {O'Brien}}]{peruzzo14}%
  \BibitemOpen
  \bibfield  {author} {\bibinfo {author} {\bibfnamefont {A.}~\bibnamefont
  {Peruzzo}}, \bibinfo {author} {\bibfnamefont {J.}~\bibnamefont {McClean}},
  \bibinfo {author} {\bibfnamefont {P.}~\bibnamefont {Shadbolt}}, \bibinfo
  {author} {\bibfnamefont {M.-H.}\ \bibnamefont {Yung}}, \bibinfo {author}
  {\bibfnamefont {X.-Q.}\ \bibnamefont {Zhou}}, \bibinfo {author}
  {\bibfnamefont {P.~J.}\ \bibnamefont {Love}}, \bibinfo {author}
  {\bibfnamefont {A.}~\bibnamefont {Aspuru-Guzik}}, \ and\ \bibinfo {author}
  {\bibfnamefont {J.~L.}\ \bibnamefont {O'Brien}},\ }\href {\doibase
  10.1038/ncomms5213} {\bibfield  {journal} {\bibinfo  {journal} {Nat.\
  Commun.}\ }\textbf {\bibinfo {volume} {5}},\ \bibinfo {pages} {4213}
  (\bibinfo {year} {2014})}\BibitemShut {NoStop}%
\bibitem [{\citenamefont {Li}\ and\ \citenamefont {Benjamin}(2017)}]{li17}%
  \BibitemOpen
  \bibfield  {author} {\bibinfo {author} {\bibfnamefont {Y.}~\bibnamefont
  {Li}}\ and\ \bibinfo {author} {\bibfnamefont {S.~C.}\ \bibnamefont
  {Benjamin}},\ }\href {\doibase 10.1103/physrevx.7.021050} {\bibfield
  {journal} {\bibinfo  {journal} {Phys.\ Rev.\ X}\ }\textbf {\bibinfo {volume}
  {7}},\ \bibinfo {pages} {021050} (\bibinfo {year} {2017})}\BibitemShut
  {NoStop}%
\bibitem [{\citenamefont {Kandala}\ \emph {et~al.}(2017)\citenamefont
  {Kandala}, \citenamefont {Mezzacapo}, \citenamefont {Temme}, \citenamefont
  {Takita}, \citenamefont {Brink}, \citenamefont {Chow},\ and\ \citenamefont
  {Gambetta}}]{Kandala.2017}%
  \BibitemOpen
  \bibfield  {author} {\bibinfo {author} {\bibfnamefont {A.}~\bibnamefont
  {Kandala}}, \bibinfo {author} {\bibfnamefont {A.}~\bibnamefont {Mezzacapo}},
  \bibinfo {author} {\bibfnamefont {K.}~\bibnamefont {Temme}}, \bibinfo
  {author} {\bibfnamefont {M.}~\bibnamefont {Takita}}, \bibinfo {author}
  {\bibfnamefont {M.}~\bibnamefont {Brink}}, \bibinfo {author} {\bibfnamefont
  {J.~M.}\ \bibnamefont {Chow}}, \ and\ \bibinfo {author} {\bibfnamefont
  {J.~M.}\ \bibnamefont {Gambetta}},\ }\href {\doibase 10.1038/nature23879}
  {\bibfield  {journal} {\bibinfo  {journal} {Nature}\ }\textbf {\bibinfo
  {volume} {549}},\ \bibinfo {pages} {242} (\bibinfo {year}
  {2017})}\BibitemShut {NoStop}%
\bibitem [{\citenamefont {Colless}\ \emph {et~al.}(2018)\citenamefont
  {Colless}, \citenamefont {Ramasesh}, \citenamefont {Dahlen}, \citenamefont
  {Blok}, \citenamefont {Kimchi-Schwartz}, \citenamefont {McClean},
  \citenamefont {Carter}, \citenamefont {Jong},\ and\ \citenamefont
  {Siddiqi}}]{colless18}%
  \BibitemOpen
  \bibfield  {author} {\bibinfo {author} {\bibfnamefont {J.~I.}\ \bibnamefont
  {Colless}}, \bibinfo {author} {\bibfnamefont {V.~V.}\ \bibnamefont
  {Ramasesh}}, \bibinfo {author} {\bibfnamefont {D.}~\bibnamefont {Dahlen}},
  \bibinfo {author} {\bibfnamefont {M.~S.}\ \bibnamefont {Blok}}, \bibinfo
  {author} {\bibfnamefont {M.~E.}\ \bibnamefont {Kimchi-Schwartz}}, \bibinfo
  {author} {\bibfnamefont {J.~R.}\ \bibnamefont {McClean}}, \bibinfo {author}
  {\bibfnamefont {J.}~\bibnamefont {Carter}}, \bibinfo {author} {\bibfnamefont
  {W.~A.~d.}\ \bibnamefont {Jong}}, \ and\ \bibinfo {author} {\bibfnamefont
  {I.}~\bibnamefont {Siddiqi}},\ }\href {\doibase 10.1103/physrevx.8.011021}
  {\bibfield  {journal} {\bibinfo  {journal} {Phys.\ Rev.\ X}\ }\textbf
  {\bibinfo {volume} {8}},\ \bibinfo {pages} {011021} (\bibinfo {year}
  {2018})}\BibitemShut {NoStop}%
\bibitem [{\citenamefont {Hempel}\ \emph {et~al.}(2018)\citenamefont {Hempel},
  \citenamefont {Maier}, \citenamefont {Romero}, \citenamefont {McClean},
  \citenamefont {Monz}, \citenamefont {Shen}, \citenamefont {Jurcevic},
  \citenamefont {Lanyon}, \citenamefont {Love}, \citenamefont {Babbush},
  \citenamefont {Aspuru-Guzik}, \citenamefont {Blatt},\ and\ \citenamefont
  {Roos}}]{hempel18}%
  \BibitemOpen
  \bibfield  {author} {\bibinfo {author} {\bibfnamefont {C.}~\bibnamefont
  {Hempel}}, \bibinfo {author} {\bibfnamefont {C.}~\bibnamefont {Maier}},
  \bibinfo {author} {\bibfnamefont {J.}~\bibnamefont {Romero}}, \bibinfo
  {author} {\bibfnamefont {J.}~\bibnamefont {McClean}}, \bibinfo {author}
  {\bibfnamefont {T.}~\bibnamefont {Monz}}, \bibinfo {author} {\bibfnamefont
  {H.}~\bibnamefont {Shen}}, \bibinfo {author} {\bibfnamefont {P.}~\bibnamefont
  {Jurcevic}}, \bibinfo {author} {\bibfnamefont {B.~P.}\ \bibnamefont
  {Lanyon}}, \bibinfo {author} {\bibfnamefont {P.}~\bibnamefont {Love}},
  \bibinfo {author} {\bibfnamefont {R.}~\bibnamefont {Babbush}}, \bibinfo
  {author} {\bibfnamefont {A.}~\bibnamefont {Aspuru-Guzik}}, \bibinfo {author}
  {\bibfnamefont {R.}~\bibnamefont {Blatt}}, \ and\ \bibinfo {author}
  {\bibfnamefont {C.~F.}\ \bibnamefont {Roos}},\ }\href {\doibase
  10.1103/physrevx.8.031022} {\bibfield  {journal} {\bibinfo  {journal} {Phys.\
  Rev.\ X}\ }\textbf {\bibinfo {volume} {8}},\ \bibinfo {pages} {031022}
  (\bibinfo {year} {2018})}\BibitemShut {NoStop}%
\bibitem [{\citenamefont {Sawaya}\ and\ \citenamefont {Huh}(2019)}]{sawaya19}%
  \BibitemOpen
  \bibfield  {author} {\bibinfo {author} {\bibfnamefont {N.~P.~D.}\
  \bibnamefont {Sawaya}}\ and\ \bibinfo {author} {\bibfnamefont
  {J.}~\bibnamefont {Huh}},\ }\href {\doibase 10.1021/acs.jpclett.9b01117}
  {\bibfield  {journal} {\bibinfo  {journal} {J.\ Phys.\ Chem.\ Lett.}\
  }\textbf {\bibinfo {volume} {10}},\ \bibinfo {pages} {3586} (\bibinfo {year}
  {2019})}\BibitemShut {NoStop}%
\bibitem [{\citenamefont {Nam}\ \emph {et~al.}(2020)\citenamefont {Nam},
  \citenamefont {Chen}, \citenamefont {Pisenti}, \citenamefont {Wright},
  \citenamefont {Delaney}, \citenamefont {Maslov}, \citenamefont {Brown},
  \citenamefont {Allen}, \citenamefont {Amini}, \citenamefont {Apisdorf},
  \citenamefont {Beck}, \citenamefont {Blinov}, \citenamefont {Chaplin},
  \citenamefont {Chmielewski}, \citenamefont {Collins}, \citenamefont
  {Debnath}, \citenamefont {Hudek}, \citenamefont {Ducore}, \citenamefont
  {Keesan}, \citenamefont {Kreikemeier}, \citenamefont {Mizrahi}, \citenamefont
  {Solomon}, \citenamefont {Williams}, \citenamefont {Wong-Campos},
  \citenamefont {Moehring}, \citenamefont {Monroe},\ and\ \citenamefont
  {Kim}}]{nam20}%
  \BibitemOpen
  \bibfield  {author} {\bibinfo {author} {\bibfnamefont {Y.}~\bibnamefont
  {Nam}}, \bibinfo {author} {\bibfnamefont {J.-S.}\ \bibnamefont {Chen}},
  \bibinfo {author} {\bibfnamefont {N.~C.}\ \bibnamefont {Pisenti}}, \bibinfo
  {author} {\bibfnamefont {K.}~\bibnamefont {Wright}}, \bibinfo {author}
  {\bibfnamefont {C.}~\bibnamefont {Delaney}}, \bibinfo {author} {\bibfnamefont
  {D.}~\bibnamefont {Maslov}}, \bibinfo {author} {\bibfnamefont {K.~R.}\
  \bibnamefont {Brown}}, \bibinfo {author} {\bibfnamefont {S.}~\bibnamefont
  {Allen}}, \bibinfo {author} {\bibfnamefont {J.~M.}\ \bibnamefont {Amini}},
  \bibinfo {author} {\bibfnamefont {J.}~\bibnamefont {Apisdorf}}, \bibinfo
  {author} {\bibfnamefont {K.~M.}\ \bibnamefont {Beck}}, \bibinfo {author}
  {\bibfnamefont {A.}~\bibnamefont {Blinov}}, \bibinfo {author} {\bibfnamefont
  {V.}~\bibnamefont {Chaplin}}, \bibinfo {author} {\bibfnamefont
  {M.}~\bibnamefont {Chmielewski}}, \bibinfo {author} {\bibfnamefont
  {C.}~\bibnamefont {Collins}}, \bibinfo {author} {\bibfnamefont
  {S.}~\bibnamefont {Debnath}}, \bibinfo {author} {\bibfnamefont {K.~M.}\
  \bibnamefont {Hudek}}, \bibinfo {author} {\bibfnamefont {A.~M.}\ \bibnamefont
  {Ducore}}, \bibinfo {author} {\bibfnamefont {M.}~\bibnamefont {Keesan}},
  \bibinfo {author} {\bibfnamefont {S.~M.}\ \bibnamefont {Kreikemeier}},
  \bibinfo {author} {\bibfnamefont {J.}~\bibnamefont {Mizrahi}}, \bibinfo
  {author} {\bibfnamefont {P.}~\bibnamefont {Solomon}}, \bibinfo {author}
  {\bibfnamefont {M.}~\bibnamefont {Williams}}, \bibinfo {author}
  {\bibfnamefont {J.~D.}\ \bibnamefont {Wong-Campos}}, \bibinfo {author}
  {\bibfnamefont {D.}~\bibnamefont {Moehring}}, \bibinfo {author}
  {\bibfnamefont {C.}~\bibnamefont {Monroe}}, \ and\ \bibinfo {author}
  {\bibfnamefont {J.}~\bibnamefont {Kim}},\ }\href {\doibase
  10.1038/s41534-020-0259-3} {\bibfield  {journal} {\bibinfo  {journal} {npj
  Quantum Inf.}\ }\textbf {\bibinfo {volume} {6}},\ \bibinfo {pages} {33}
  (\bibinfo {year} {2020})}\BibitemShut {NoStop}%
\bibitem [{\citenamefont {Arute}\ \emph {et~al.}(2020)\citenamefont {Arute},
  \citenamefont {Arya}, \citenamefont {Babbush}, \citenamefont {Bacon},
  \citenamefont {Bardin}, \citenamefont {Barends}, \citenamefont {Boixo},
  \citenamefont {Broughton}, \citenamefont {Buckley}, \citenamefont {Buell},
  \citenamefont {Burkett}, \citenamefont {Bushnell}, \citenamefont {Chen},
  \citenamefont {Chen}, \citenamefont {Chiaro}, \citenamefont {Collins},
  \citenamefont {Courtney}, \citenamefont {Demura}, \citenamefont {Dunsworth},
  \citenamefont {Eppens}, \citenamefont {Farhi}, \citenamefont {Fowler},
  \citenamefont {Foxen}, \citenamefont {Gidney}, \citenamefont {Giustina},
  \citenamefont {Graff}, \citenamefont {Habegger}, \citenamefont {Harrigan},
  \citenamefont {Ho}, \citenamefont {Hong}, \citenamefont {Huang},
  \citenamefont {Huggins}, \citenamefont {Ioffe}, \citenamefont {Isakov},
  \citenamefont {Jeffrey}, \citenamefont {Jiang}, \citenamefont {Jones},
  \citenamefont {Kafri}, \citenamefont {Kechedzhi}, \citenamefont {Kelly},
  \citenamefont {Kim}, \citenamefont {Klimov}, \citenamefont {Korotkov},
  \citenamefont {Kostritsa}, \citenamefont {Landhuis}, \citenamefont {Laptev},
  \citenamefont {Lindmark}, \citenamefont {Lucero}, \citenamefont {Martin},
  \citenamefont {Martinis}, \citenamefont {McClean}, \citenamefont {McEwen},
  \citenamefont {Megrant}, \citenamefont {Mi}, \citenamefont {Mohseni},
  \citenamefont {Mruczkiewicz}, \citenamefont {Mutus}, \citenamefont {Naaman},
  \citenamefont {Neeley}, \citenamefont {Neill}, \citenamefont {Neven},
  \citenamefont {Niu}, \citenamefont {O'Brien}, \citenamefont {Ostby},
  \citenamefont {Petukhov}, \citenamefont {Putterman}, \citenamefont
  {Quintana}, \citenamefont {Roushan}, \citenamefont {Rubin}, \citenamefont
  {Sank}, \citenamefont {Satzinger}, \citenamefont {Smelyanskiy}, \citenamefont
  {Strain}, \citenamefont {Sung}, \citenamefont {Szalay}, \citenamefont
  {Takeshita}, \citenamefont {Vainsencher}, \citenamefont {White},
  \citenamefont {Wiebe}, \citenamefont {Yao}, \citenamefont {Yeh},\ and\
  \citenamefont {Zalcman}}]{rubin20}%
  \BibitemOpen
  \bibfield  {author} {\bibinfo {author} {\bibfnamefont {F.}~\bibnamefont
  {Arute}}, \bibinfo {author} {\bibfnamefont {K.}~\bibnamefont {Arya}},
  \bibinfo {author} {\bibfnamefont {R.}~\bibnamefont {Babbush}}, \bibinfo
  {author} {\bibfnamefont {D.}~\bibnamefont {Bacon}}, \bibinfo {author}
  {\bibfnamefont {J.~C.}\ \bibnamefont {Bardin}}, \bibinfo {author}
  {\bibfnamefont {R.}~\bibnamefont {Barends}}, \bibinfo {author} {\bibfnamefont
  {S.}~\bibnamefont {Boixo}}, \bibinfo {author} {\bibfnamefont
  {M.}~\bibnamefont {Broughton}}, \bibinfo {author} {\bibfnamefont {B.~B.}\
  \bibnamefont {Buckley}}, \bibinfo {author} {\bibfnamefont {D.~A.}\
  \bibnamefont {Buell}}, \bibinfo {author} {\bibfnamefont {B.}~\bibnamefont
  {Burkett}}, \bibinfo {author} {\bibfnamefont {N.}~\bibnamefont {Bushnell}},
  \bibinfo {author} {\bibfnamefont {Y.}~\bibnamefont {Chen}}, \bibinfo {author}
  {\bibfnamefont {Z.}~\bibnamefont {Chen}}, \bibinfo {author} {\bibfnamefont
  {B.}~\bibnamefont {Chiaro}}, \bibinfo {author} {\bibfnamefont
  {R.}~\bibnamefont {Collins}}, \bibinfo {author} {\bibfnamefont
  {W.}~\bibnamefont {Courtney}}, \bibinfo {author} {\bibfnamefont
  {S.}~\bibnamefont {Demura}}, \bibinfo {author} {\bibfnamefont
  {A.}~\bibnamefont {Dunsworth}}, \bibinfo {author} {\bibfnamefont
  {D.}~\bibnamefont {Eppens}}, \bibinfo {author} {\bibfnamefont
  {E.}~\bibnamefont {Farhi}}, \bibinfo {author} {\bibfnamefont
  {A.}~\bibnamefont {Fowler}}, \bibinfo {author} {\bibfnamefont
  {B.}~\bibnamefont {Foxen}}, \bibinfo {author} {\bibfnamefont
  {C.}~\bibnamefont {Gidney}}, \bibinfo {author} {\bibfnamefont
  {M.}~\bibnamefont {Giustina}}, \bibinfo {author} {\bibfnamefont
  {R.}~\bibnamefont {Graff}}, \bibinfo {author} {\bibfnamefont
  {S.}~\bibnamefont {Habegger}}, \bibinfo {author} {\bibfnamefont {M.~P.}\
  \bibnamefont {Harrigan}}, \bibinfo {author} {\bibfnamefont {A.}~\bibnamefont
  {Ho}}, \bibinfo {author} {\bibfnamefont {S.}~\bibnamefont {Hong}}, \bibinfo
  {author} {\bibfnamefont {T.}~\bibnamefont {Huang}}, \bibinfo {author}
  {\bibfnamefont {W.~J.}\ \bibnamefont {Huggins}}, \bibinfo {author}
  {\bibfnamefont {L.}~\bibnamefont {Ioffe}}, \bibinfo {author} {\bibfnamefont
  {S.~V.}\ \bibnamefont {Isakov}}, \bibinfo {author} {\bibfnamefont
  {E.}~\bibnamefont {Jeffrey}}, \bibinfo {author} {\bibfnamefont
  {Z.}~\bibnamefont {Jiang}}, \bibinfo {author} {\bibfnamefont
  {C.}~\bibnamefont {Jones}}, \bibinfo {author} {\bibfnamefont
  {D.}~\bibnamefont {Kafri}}, \bibinfo {author} {\bibfnamefont
  {K.}~\bibnamefont {Kechedzhi}}, \bibinfo {author} {\bibfnamefont
  {J.}~\bibnamefont {Kelly}}, \bibinfo {author} {\bibfnamefont
  {S.}~\bibnamefont {Kim}}, \bibinfo {author} {\bibfnamefont {P.~V.}\
  \bibnamefont {Klimov}}, \bibinfo {author} {\bibfnamefont {A.}~\bibnamefont
  {Korotkov}}, \bibinfo {author} {\bibfnamefont {F.}~\bibnamefont {Kostritsa}},
  \bibinfo {author} {\bibfnamefont {D.}~\bibnamefont {Landhuis}}, \bibinfo
  {author} {\bibfnamefont {P.}~\bibnamefont {Laptev}}, \bibinfo {author}
  {\bibfnamefont {M.}~\bibnamefont {Lindmark}}, \bibinfo {author}
  {\bibfnamefont {E.}~\bibnamefont {Lucero}}, \bibinfo {author} {\bibfnamefont
  {O.}~\bibnamefont {Martin}}, \bibinfo {author} {\bibfnamefont {J.~M.}\
  \bibnamefont {Martinis}}, \bibinfo {author} {\bibfnamefont {J.~R.}\
  \bibnamefont {McClean}}, \bibinfo {author} {\bibfnamefont {M.}~\bibnamefont
  {McEwen}}, \bibinfo {author} {\bibfnamefont {A.}~\bibnamefont {Megrant}},
  \bibinfo {author} {\bibfnamefont {X.}~\bibnamefont {Mi}}, \bibinfo {author}
  {\bibfnamefont {M.}~\bibnamefont {Mohseni}}, \bibinfo {author} {\bibfnamefont
  {W.}~\bibnamefont {Mruczkiewicz}}, \bibinfo {author} {\bibfnamefont
  {J.}~\bibnamefont {Mutus}}, \bibinfo {author} {\bibfnamefont
  {O.}~\bibnamefont {Naaman}}, \bibinfo {author} {\bibfnamefont
  {M.}~\bibnamefont {Neeley}}, \bibinfo {author} {\bibfnamefont
  {C.}~\bibnamefont {Neill}}, \bibinfo {author} {\bibfnamefont
  {H.}~\bibnamefont {Neven}}, \bibinfo {author} {\bibfnamefont {M.~Y.}\
  \bibnamefont {Niu}}, \bibinfo {author} {\bibfnamefont {T.~E.}\ \bibnamefont
  {O'Brien}}, \bibinfo {author} {\bibfnamefont {E.}~\bibnamefont {Ostby}},
  \bibinfo {author} {\bibfnamefont {A.}~\bibnamefont {Petukhov}}, \bibinfo
  {author} {\bibfnamefont {H.}~\bibnamefont {Putterman}}, \bibinfo {author}
  {\bibfnamefont {C.}~\bibnamefont {Quintana}}, \bibinfo {author}
  {\bibfnamefont {P.}~\bibnamefont {Roushan}}, \bibinfo {author} {\bibfnamefont
  {N.~C.}\ \bibnamefont {Rubin}}, \bibinfo {author} {\bibfnamefont
  {D.}~\bibnamefont {Sank}}, \bibinfo {author} {\bibfnamefont {K.~J.}\
  \bibnamefont {Satzinger}}, \bibinfo {author} {\bibfnamefont {V.}~\bibnamefont
  {Smelyanskiy}}, \bibinfo {author} {\bibfnamefont {D.}~\bibnamefont {Strain}},
  \bibinfo {author} {\bibfnamefont {K.~J.}\ \bibnamefont {Sung}}, \bibinfo
  {author} {\bibfnamefont {M.}~\bibnamefont {Szalay}}, \bibinfo {author}
  {\bibfnamefont {T.~Y.}\ \bibnamefont {Takeshita}}, \bibinfo {author}
  {\bibfnamefont {A.}~\bibnamefont {Vainsencher}}, \bibinfo {author}
  {\bibfnamefont {T.}~\bibnamefont {White}}, \bibinfo {author} {\bibfnamefont
  {N.}~\bibnamefont {Wiebe}}, \bibinfo {author} {\bibfnamefont {Z.~J.}\
  \bibnamefont {Yao}}, \bibinfo {author} {\bibfnamefont {P.}~\bibnamefont
  {Yeh}}, \ and\ \bibinfo {author} {\bibfnamefont {A.}~\bibnamefont
  {Zalcman}},\ }\href {\doibase 10.1126/science.abb9811} {\bibfield  {journal}
  {\bibinfo  {journal} {Science}\ }\textbf {\bibinfo {volume} {369}},\ \bibinfo
  {pages} {1084} (\bibinfo {year} {2020})}\BibitemShut {NoStop}%
\bibitem [{\citenamefont {McArdle}\ \emph {et~al.}(2019)\citenamefont
  {McArdle}, \citenamefont {Mayorov}, \citenamefont {Shan}, \citenamefont
  {Benjamin},\ and\ \citenamefont {Yuan}}]{mcardle19}%
  \BibitemOpen
  \bibfield  {author} {\bibinfo {author} {\bibfnamefont {S.}~\bibnamefont
  {McArdle}}, \bibinfo {author} {\bibfnamefont {A.}~\bibnamefont {Mayorov}},
  \bibinfo {author} {\bibfnamefont {X.}~\bibnamefont {Shan}}, \bibinfo {author}
  {\bibfnamefont {S.}~\bibnamefont {Benjamin}}, \ and\ \bibinfo {author}
  {\bibfnamefont {X.}~\bibnamefont {Yuan}},\ }\href {\doibase
  10.1039/c9sc01313j} {\bibfield  {journal} {\bibinfo  {journal} {Chem.\ Sci.}\
  }\textbf {\bibinfo {volume} {10}},\ \bibinfo {pages} {5725} (\bibinfo {year}
  {2019})}\BibitemShut {NoStop}%
\bibitem [{\citenamefont {Sawaya}\ \emph {et~al.}(2020)\citenamefont {Sawaya},
  \citenamefont {Paesani},\ and\ \citenamefont {Tabor}}]{sawaya20}%
  \BibitemOpen
  \bibfield  {author} {\bibinfo {author} {\bibfnamefont {N.~P.~D.}\
  \bibnamefont {Sawaya}}, \bibinfo {author} {\bibfnamefont {F.}~\bibnamefont
  {Paesani}}, \ and\ \bibinfo {author} {\bibfnamefont {D.~P.}\ \bibnamefont
  {Tabor}},\ }\href@noop {} {\bibfield  {journal} {\bibinfo  {journal} {arXiv}\
  } (\bibinfo {year} {2020})},\ \Eprint {http://arxiv.org/abs/2009.05066}
  {2009.05066} \BibitemShut {NoStop}%
\bibitem [{\citenamefont {Kassal}\ \emph {et~al.}(2008)\citenamefont {Kassal},
  \citenamefont {Jordan}, \citenamefont {Love}, \citenamefont {Mohseni},\ and\
  \citenamefont {Aspuru-Guzik}}]{kassal08}%
  \BibitemOpen
  \bibfield  {author} {\bibinfo {author} {\bibfnamefont {I.}~\bibnamefont
  {Kassal}}, \bibinfo {author} {\bibfnamefont {S.~P.}\ \bibnamefont {Jordan}},
  \bibinfo {author} {\bibfnamefont {P.~J.}\ \bibnamefont {Love}}, \bibinfo
  {author} {\bibfnamefont {M.}~\bibnamefont {Mohseni}}, \ and\ \bibinfo
  {author} {\bibfnamefont {A.}~\bibnamefont {Aspuru-Guzik}},\ }\href {\doibase
  10.1073/pnas.0808245105} {\bibfield  {journal} {\bibinfo  {journal} {Proc.\
  Natl.\ Acad.\ Sci.}\ }\textbf {\bibinfo {volume} {105}},\ \bibinfo {pages}
  {18681} (\bibinfo {year} {2008})}\BibitemShut {NoStop}%
\bibitem [{\citenamefont {Ollitrault}\ \emph {et~al.}(2020)\citenamefont
  {Ollitrault}, \citenamefont {Mazzola},\ and\ \citenamefont
  {Tavernelli}}]{ollitrault20b}%
  \BibitemOpen
  \bibfield  {author} {\bibinfo {author} {\bibfnamefont {P.~J.}\ \bibnamefont
  {Ollitrault}}, \bibinfo {author} {\bibfnamefont {G.}~\bibnamefont {Mazzola}},
  \ and\ \bibinfo {author} {\bibfnamefont {I.}~\bibnamefont {Tavernelli}},\
  }\href {\doibase 10.1103/PhysRevLett.125.260511} {\bibfield  {journal}
  {\bibinfo  {journal} {Phys.\ Rev.\ Lett.}\ }\textbf {\bibinfo {volume}
  {125}},\ \bibinfo {pages} {260511} (\bibinfo {year} {2020})}\BibitemShut
  {NoStop}%
\bibitem [{\citenamefont {Britton}\ \emph {et~al.}(2012)\citenamefont
  {Britton}, \citenamefont {Sawyer}, \citenamefont {Keith}, \citenamefont
  {Wang}, \citenamefont {Freericks}, \citenamefont {Uys}, \citenamefont
  {Biercuk},\ and\ \citenamefont {Bollinger}}]{britton12}%
  \BibitemOpen
  \bibfield  {author} {\bibinfo {author} {\bibfnamefont {J.~W.}\ \bibnamefont
  {Britton}}, \bibinfo {author} {\bibfnamefont {B.~C.}\ \bibnamefont {Sawyer}},
  \bibinfo {author} {\bibfnamefont {A.~C.}\ \bibnamefont {Keith}}, \bibinfo
  {author} {\bibfnamefont {C.-C.~J.}\ \bibnamefont {Wang}}, \bibinfo {author}
  {\bibfnamefont {J.~K.}\ \bibnamefont {Freericks}}, \bibinfo {author}
  {\bibfnamefont {H.}~\bibnamefont {Uys}}, \bibinfo {author} {\bibfnamefont
  {M.~J.}\ \bibnamefont {Biercuk}}, \ and\ \bibinfo {author} {\bibfnamefont
  {J.~J.}\ \bibnamefont {Bollinger}},\ }\href {\doibase 10.1038/nature10981}
  {\bibfield  {journal} {\bibinfo  {journal} {Nature}\ }\textbf {\bibinfo
  {volume} {484}},\ \bibinfo {pages} {489} (\bibinfo {year}
  {2012})}\BibitemShut {NoStop}%
\bibitem [{\citenamefont {Zhang}\ \emph {et~al.}(2017)\citenamefont {Zhang},
  \citenamefont {Pagano}, \citenamefont {Hess}, \citenamefont {Kyprianidis},
  \citenamefont {Becker}, \citenamefont {Kaplan}, \citenamefont {Gorshkov},
  \citenamefont {Gong},\ and\ \citenamefont {Monroe}}]{zhang17}%
  \BibitemOpen
  \bibfield  {author} {\bibinfo {author} {\bibfnamefont {J.}~\bibnamefont
  {Zhang}}, \bibinfo {author} {\bibfnamefont {G.}~\bibnamefont {Pagano}},
  \bibinfo {author} {\bibfnamefont {P.~W.}\ \bibnamefont {Hess}}, \bibinfo
  {author} {\bibfnamefont {A.}~\bibnamefont {Kyprianidis}}, \bibinfo {author}
  {\bibfnamefont {P.}~\bibnamefont {Becker}}, \bibinfo {author} {\bibfnamefont
  {H.}~\bibnamefont {Kaplan}}, \bibinfo {author} {\bibfnamefont {A.~V.}\
  \bibnamefont {Gorshkov}}, \bibinfo {author} {\bibfnamefont {Z.-X.}\
  \bibnamefont {Gong}}, \ and\ \bibinfo {author} {\bibfnamefont
  {C.}~\bibnamefont {Monroe}},\ }\href {\doibase 10.1038/nature24654}
  {\bibfield  {journal} {\bibinfo  {journal} {Nature}\ }\textbf {\bibinfo
  {volume} {551}},\ \bibinfo {pages} {601} (\bibinfo {year}
  {2017})}\BibitemShut {NoStop}%
\bibitem [{\citenamefont {Arg{\" u}ello-Luengo}\ \emph
  {et~al.}(2019)\citenamefont {Arg{\" u}ello-Luengo}, \citenamefont {Gonz{\'
  a}lez-Tudela}, \citenamefont {Shi}, \citenamefont {Zoller},\ and\
  \citenamefont {Cirac}}]{arguelloluengo19}%
  \BibitemOpen
  \bibfield  {author} {\bibinfo {author} {\bibfnamefont {J.}~\bibnamefont
  {Arg{\" u}ello-Luengo}}, \bibinfo {author} {\bibfnamefont {A.}~\bibnamefont
  {Gonz{\' a}lez-Tudela}}, \bibinfo {author} {\bibfnamefont {T.}~\bibnamefont
  {Shi}}, \bibinfo {author} {\bibfnamefont {P.}~\bibnamefont {Zoller}}, \ and\
  \bibinfo {author} {\bibfnamefont {J.~I.}\ \bibnamefont {Cirac}},\ }\href
  {\doibase 10.1038/s41586-019-1614-4} {\bibfield  {journal} {\bibinfo
  {journal} {Nature}\ }\textbf {\bibinfo {volume} {574}},\ \bibinfo {pages}
  {215} (\bibinfo {year} {2019})}\BibitemShut {NoStop}%
\bibitem [{\citenamefont {Sparrow}\ \emph {et~al.}(2018)\citenamefont
  {Sparrow}, \citenamefont {Mart{\' i}n-L{\' o}pez}, \citenamefont
  {Maraviglia}, \citenamefont {Neville}, \citenamefont {Harrold}, \citenamefont
  {Carolan}, \citenamefont {Joglekar}, \citenamefont {Hashimoto}, \citenamefont
  {Matsuda}, \citenamefont {O'Brien}, \citenamefont {Tew},\ and\ \citenamefont
  {Laing}}]{sparrow18}%
  \BibitemOpen
  \bibfield  {author} {\bibinfo {author} {\bibfnamefont {C.}~\bibnamefont
  {Sparrow}}, \bibinfo {author} {\bibfnamefont {E.}~\bibnamefont {Mart{\'
  i}n-L{\' o}pez}}, \bibinfo {author} {\bibfnamefont {N.}~\bibnamefont
  {Maraviglia}}, \bibinfo {author} {\bibfnamefont {A.}~\bibnamefont {Neville}},
  \bibinfo {author} {\bibfnamefont {C.}~\bibnamefont {Harrold}}, \bibinfo
  {author} {\bibfnamefont {J.}~\bibnamefont {Carolan}}, \bibinfo {author}
  {\bibfnamefont {Y.~N.}\ \bibnamefont {Joglekar}}, \bibinfo {author}
  {\bibfnamefont {T.}~\bibnamefont {Hashimoto}}, \bibinfo {author}
  {\bibfnamefont {N.}~\bibnamefont {Matsuda}}, \bibinfo {author} {\bibfnamefont
  {J.~L.}\ \bibnamefont {O'Brien}}, \bibinfo {author} {\bibfnamefont {D.~P.}\
  \bibnamefont {Tew}}, \ and\ \bibinfo {author} {\bibfnamefont
  {A.}~\bibnamefont {Laing}},\ }\href {\doibase 10.1038/s41586-018-0152-9}
  {\bibfield  {journal} {\bibinfo  {journal} {Nature}\ }\textbf {\bibinfo
  {volume} {557}},\ \bibinfo {pages} {660} (\bibinfo {year}
  {2018})}\BibitemShut {NoStop}%
\bibitem [{\citenamefont {Chen}\ \emph {et~al.}(2021)\citenamefont {Chen},
  \citenamefont {Gan}, \citenamefont {Zhang}, \citenamefont {Matuskevich},\
  and\ \citenamefont {Kim}}]{chen21}%
  \BibitemOpen
  \bibfield  {author} {\bibinfo {author} {\bibfnamefont {W.}~\bibnamefont
  {Chen}}, \bibinfo {author} {\bibfnamefont {J.}~\bibnamefont {Gan}}, \bibinfo
  {author} {\bibfnamefont {J.-N.}\ \bibnamefont {Zhang}}, \bibinfo {author}
  {\bibfnamefont {D.}~\bibnamefont {Matuskevich}}, \ and\ \bibinfo {author}
  {\bibfnamefont {K.}~\bibnamefont {Kim}},\ }\href@noop {} {\bibfield
  {journal} {\bibinfo  {journal} {arXiv}\ } (\bibinfo {year} {2021})},\ \Eprint
  {http://arxiv.org/abs/2103.14299} {2103.14299} \BibitemShut {NoStop}%
\bibitem [{\citenamefont {Huh}\ \emph {et~al.}(2015)\citenamefont {Huh},
  \citenamefont {Guerreschi}, \citenamefont {Peropadre}, \citenamefont
  {McClean},\ and\ \citenamefont {Aspuru-Guzik}}]{huh15}%
  \BibitemOpen
  \bibfield  {author} {\bibinfo {author} {\bibfnamefont {J.}~\bibnamefont
  {Huh}}, \bibinfo {author} {\bibfnamefont {G.~G.}\ \bibnamefont {Guerreschi}},
  \bibinfo {author} {\bibfnamefont {B.}~\bibnamefont {Peropadre}}, \bibinfo
  {author} {\bibfnamefont {J.~R.}\ \bibnamefont {McClean}}, \ and\ \bibinfo
  {author} {\bibfnamefont {A.}~\bibnamefont {Aspuru-Guzik}},\ }\href {\doibase
  10.1038/nphoton.2015.153} {\bibfield  {journal} {\bibinfo  {journal} {Nat.\
  Photonics}\ }\textbf {\bibinfo {volume} {9}},\ \bibinfo {pages} {615}
  (\bibinfo {year} {2015})}\BibitemShut {NoStop}%
\bibitem [{\citenamefont {Wang}\ \emph {et~al.}(2020)\citenamefont {Wang},
  \citenamefont {Curtis}, \citenamefont {Lester}, \citenamefont {Zhang},
  \citenamefont {Gao}, \citenamefont {Freeze}, \citenamefont {Batista},
  \citenamefont {Vaccaro}, \citenamefont {Chuang}, \citenamefont {Frunzio},
  \citenamefont {Jiang}, \citenamefont {Girvin},\ and\ \citenamefont
  {Schoelkopf}}]{wang20}%
  \BibitemOpen
  \bibfield  {author} {\bibinfo {author} {\bibfnamefont {C.~S.}\ \bibnamefont
  {Wang}}, \bibinfo {author} {\bibfnamefont {J.~C.}\ \bibnamefont {Curtis}},
  \bibinfo {author} {\bibfnamefont {B.~J.}\ \bibnamefont {Lester}}, \bibinfo
  {author} {\bibfnamefont {Y.}~\bibnamefont {Zhang}}, \bibinfo {author}
  {\bibfnamefont {Y.~Y.}\ \bibnamefont {Gao}}, \bibinfo {author} {\bibfnamefont
  {J.}~\bibnamefont {Freeze}}, \bibinfo {author} {\bibfnamefont {V.~S.}\
  \bibnamefont {Batista}}, \bibinfo {author} {\bibfnamefont {P.~H.}\
  \bibnamefont {Vaccaro}}, \bibinfo {author} {\bibfnamefont {I.~L.}\
  \bibnamefont {Chuang}}, \bibinfo {author} {\bibfnamefont {L.}~\bibnamefont
  {Frunzio}}, \bibinfo {author} {\bibfnamefont {L.}~\bibnamefont {Jiang}},
  \bibinfo {author} {\bibfnamefont {S.~M.}\ \bibnamefont {Girvin}}, \ and\
  \bibinfo {author} {\bibfnamefont {R.~J.}\ \bibnamefont {Schoelkopf}},\ }\href
  {\doibase 10.1103/physrevx.10.021060} {\bibfield  {journal} {\bibinfo
  {journal} {Phys.\ Rev.\ X}\ }\textbf {\bibinfo {volume} {10}},\ \bibinfo
  {pages} {021060} (\bibinfo {year} {2020})}\BibitemShut {NoStop}%
\bibitem [{\citenamefont {Jnane}\ \emph {et~al.}(2020)\citenamefont {Jnane},
  \citenamefont {Sawaya}, \citenamefont {Peropadre}, \citenamefont
  {Aspuru-Guzik}, \citenamefont {Garcia-Patron},\ and\ \citenamefont
  {Huh}}]{jnane20}%
  \BibitemOpen
  \bibfield  {author} {\bibinfo {author} {\bibfnamefont {H.}~\bibnamefont
  {Jnane}}, \bibinfo {author} {\bibfnamefont {N.~P.~D.}\ \bibnamefont
  {Sawaya}}, \bibinfo {author} {\bibfnamefont {B.}~\bibnamefont {Peropadre}},
  \bibinfo {author} {\bibfnamefont {A.}~\bibnamefont {Aspuru-Guzik}}, \bibinfo
  {author} {\bibfnamefont {R.}~\bibnamefont {Garcia-Patron}}, \ and\ \bibinfo
  {author} {\bibfnamefont {J.}~\bibnamefont {Huh}},\ }\href@noop {} {\bibfield
  {journal} {\bibinfo  {journal} {arXiv}\ } (\bibinfo {year} {2020})},\ \Eprint
  {http://arxiv.org/abs/2011.05553} {2011.05553} \BibitemShut {NoStop}%
\bibitem [{\citenamefont {Casanova}\ \emph {et~al.}(2012)\citenamefont
  {Casanova}, \citenamefont {Mezzacapo}, \citenamefont {Lamata},\ and\
  \citenamefont {Solano}}]{casanova12}%
  \BibitemOpen
  \bibfield  {author} {\bibinfo {author} {\bibfnamefont {J.}~\bibnamefont
  {Casanova}}, \bibinfo {author} {\bibfnamefont {A.}~\bibnamefont {Mezzacapo}},
  \bibinfo {author} {\bibfnamefont {L.}~\bibnamefont {Lamata}}, \ and\ \bibinfo
  {author} {\bibfnamefont {E.}~\bibnamefont {Solano}},\ }\href {\doibase
  10.1103/PhysRevLett.108.190502} {\bibfield  {journal} {\bibinfo  {journal}
  {Phys.\ Rev.\ Lett.}\ }\textbf {\bibinfo {volume} {108}},\ \bibinfo {pages}
  {190502} (\bibinfo {year} {2012})}\BibitemShut {NoStop}%
\bibitem [{\citenamefont {Lamata}\ \emph {et~al.}(2014)\citenamefont {Lamata},
  \citenamefont {Mezzacapo}, \citenamefont {Casanova},\ and\ \citenamefont
  {Solano}}]{lamata14}%
  \BibitemOpen
  \bibfield  {author} {\bibinfo {author} {\bibfnamefont {L.}~\bibnamefont
  {Lamata}}, \bibinfo {author} {\bibfnamefont {A.}~\bibnamefont {Mezzacapo}},
  \bibinfo {author} {\bibfnamefont {J.}~\bibnamefont {Casanova}}, \ and\
  \bibinfo {author} {\bibfnamefont {E.}~\bibnamefont {Solano}},\ }\href
  {\doibase 10.1140/epjqt9} {\bibfield  {journal} {\bibinfo  {journal} {EPJ
  Quantum Technol.}\ }\textbf {\bibinfo {volume} {1}},\ \bibinfo {pages} {9}
  (\bibinfo {year} {2014})}\BibitemShut {NoStop}%
\bibitem [{\citenamefont {Lamata}\ \emph {et~al.}(2018)\citenamefont {Lamata},
  \citenamefont {Parra-Rodriguez}, \citenamefont {Sanz},\ and\ \citenamefont
  {Solano}}]{lamata18}%
  \BibitemOpen
  \bibfield  {author} {\bibinfo {author} {\bibfnamefont {L.}~\bibnamefont
  {Lamata}}, \bibinfo {author} {\bibfnamefont {A.}~\bibnamefont
  {Parra-Rodriguez}}, \bibinfo {author} {\bibfnamefont {M.}~\bibnamefont
  {Sanz}}, \ and\ \bibinfo {author} {\bibfnamefont {E.}~\bibnamefont
  {Solano}},\ }\href {\doibase 10.1080/23746149.2018.1457981} {\bibfield
  {journal} {\bibinfo  {journal} {Adv.\ Phys.\ X}\ }\textbf {\bibinfo {volume}
  {3}},\ \bibinfo {pages} {1457981} (\bibinfo {year} {2018})}\BibitemShut
  {NoStop}%
\bibitem [{\citenamefont {Garc{\' i}a-{\' A}lvarez}\ \emph
  {et~al.}(2015)\citenamefont {Garc{\' i}a-{\' A}lvarez}, \citenamefont {{Las
  Heras}}, \citenamefont {Mezzacapo}, \citenamefont {Sanz}, \citenamefont
  {Solano},\ and\ \citenamefont {Lamata}}]{garciaalvarez15}%
  \BibitemOpen
  \bibfield  {author} {\bibinfo {author} {\bibfnamefont {L.}~\bibnamefont
  {Garc{\' i}a-{\' A}lvarez}}, \bibinfo {author} {\bibfnamefont
  {U.}~\bibnamefont {{Las Heras}}}, \bibinfo {author} {\bibfnamefont
  {A.}~\bibnamefont {Mezzacapo}}, \bibinfo {author} {\bibfnamefont
  {M.}~\bibnamefont {Sanz}}, \bibinfo {author} {\bibfnamefont {E.}~\bibnamefont
  {Solano}}, \ and\ \bibinfo {author} {\bibfnamefont {L.}~\bibnamefont
  {Lamata}},\ }\href {\doibase 10.1038/srep27836} {\bibfield  {journal}
  {\bibinfo  {journal} {Sci.\ Rep.}\ }\textbf {\bibinfo {volume} {6}},\
  \bibinfo {pages} {27836} (\bibinfo {year} {2015})}\BibitemShut {NoStop}%
\bibitem [{\citenamefont {Gorman}\ \emph {et~al.}(2018)\citenamefont {Gorman},
  \citenamefont {Hemmerling}, \citenamefont {Megidish}, \citenamefont
  {Moeller}, \citenamefont {Schindler}, \citenamefont {Sarovar},\ and\
  \citenamefont {Haeffner}}]{gorman18}%
  \BibitemOpen
  \bibfield  {author} {\bibinfo {author} {\bibfnamefont {D.~J.}\ \bibnamefont
  {Gorman}}, \bibinfo {author} {\bibfnamefont {B.}~\bibnamefont {Hemmerling}},
  \bibinfo {author} {\bibfnamefont {E.}~\bibnamefont {Megidish}}, \bibinfo
  {author} {\bibfnamefont {S.~A.}\ \bibnamefont {Moeller}}, \bibinfo {author}
  {\bibfnamefont {P.}~\bibnamefont {Schindler}}, \bibinfo {author}
  {\bibfnamefont {M.}~\bibnamefont {Sarovar}}, \ and\ \bibinfo {author}
  {\bibfnamefont {H.}~\bibnamefont {Haeffner}},\ }\href {\doibase
  10.1103/physrevx.8.011038} {\bibfield  {journal} {\bibinfo  {journal} {Phys.\
  Rev.\ X}\ }\textbf {\bibinfo {volume} {8}},\ \bibinfo {pages} {011038}
  (\bibinfo {year} {2018})}\BibitemShut {NoStop}%
\bibitem [{\citenamefont {Lemmer}\ \emph {et~al.}(2018)\citenamefont {Lemmer},
  \citenamefont {Cormick}, \citenamefont {Tamascelli}, \citenamefont {Schaetz},
  \citenamefont {Huelga},\ and\ \citenamefont {Plenio}}]{lemmer18}%
  \BibitemOpen
  \bibfield  {author} {\bibinfo {author} {\bibfnamefont {A.}~\bibnamefont
  {Lemmer}}, \bibinfo {author} {\bibfnamefont {C.}~\bibnamefont {Cormick}},
  \bibinfo {author} {\bibfnamefont {D.}~\bibnamefont {Tamascelli}}, \bibinfo
  {author} {\bibfnamefont {T.}~\bibnamefont {Schaetz}}, \bibinfo {author}
  {\bibfnamefont {S.~F.}\ \bibnamefont {Huelga}}, \ and\ \bibinfo {author}
  {\bibfnamefont {M.~B.}\ \bibnamefont {Plenio}},\ }\href {\doibase
  10.1088/1367-2630/aac87d} {\bibfield  {journal} {\bibinfo  {journal} {New J.\
  Phys.}\ }\textbf {\bibinfo {volume} {20}},\ \bibinfo {pages} {073002}
  (\bibinfo {year} {2018})}\BibitemShut {NoStop}%
\bibitem [{\citenamefont {Schlawin}\ \emph {et~al.}(2020)\citenamefont
  {Schlawin}, \citenamefont {Gessner}, \citenamefont {Buchleitner},
  \citenamefont {Schaetz},\ and\ \citenamefont {Skourtis}}]{schlawin20}%
  \BibitemOpen
  \bibfield  {author} {\bibinfo {author} {\bibfnamefont {F.}~\bibnamefont
  {Schlawin}}, \bibinfo {author} {\bibfnamefont {M.}~\bibnamefont {Gessner}},
  \bibinfo {author} {\bibfnamefont {A.}~\bibnamefont {Buchleitner}}, \bibinfo
  {author} {\bibfnamefont {T.}~\bibnamefont {Schaetz}}, \ and\ \bibinfo
  {author} {\bibfnamefont {S.~S.}\ \bibnamefont {Skourtis}},\ }\href@noop {}
  {\bibfield  {journal} {\bibinfo  {journal} {arXiv}\ } (\bibinfo {year}
  {2020})},\ \Eprint {http://arxiv.org/abs/2004.02925} {2004.02925}
  \BibitemShut {NoStop}%
\bibitem [{\citenamefont {Seidner}\ \emph {et~al.}(1992)\citenamefont
  {Seidner}, \citenamefont {Stock}, \citenamefont {Sobolewski},\ and\
  \citenamefont {Domcke}}]{seidner92}%
  \BibitemOpen
  \bibfield  {author} {\bibinfo {author} {\bibfnamefont {L.}~\bibnamefont
  {Seidner}}, \bibinfo {author} {\bibfnamefont {G.}~\bibnamefont {Stock}},
  \bibinfo {author} {\bibfnamefont {A.~L.}\ \bibnamefont {Sobolewski}}, \ and\
  \bibinfo {author} {\bibfnamefont {W.}~\bibnamefont {Domcke}},\ }\href
  {\doibase 10.1063/1.462715} {\bibfield  {journal} {\bibinfo  {journal} {J.\
  Chem.\ Phys.}\ }\textbf {\bibinfo {volume} {96}},\ \bibinfo {pages} {5298}
  (\bibinfo {year} {1992})}\BibitemShut {NoStop}%
\bibitem [{\citenamefont {M{\o}lmer}\ and\ \citenamefont
  {S{\o}rensen}(1999)}]{molmer99}%
  \BibitemOpen
  \bibfield  {author} {\bibinfo {author} {\bibfnamefont {K.}~\bibnamefont
  {M{\o}lmer}}\ and\ \bibinfo {author} {\bibfnamefont {A.}~\bibnamefont
  {S{\o}rensen}},\ }\href {\doibase 10.1103/PhysRevLett.82.1835} {\bibfield
  {journal} {\bibinfo  {journal} {Phys.\ Rev.\ Lett.}\ }\textbf {\bibinfo
  {volume} {82}},\ \bibinfo {pages} {1835} (\bibinfo {year}
  {1999})}\BibitemShut {NoStop}%
\bibitem [{\citenamefont {Leibfried}\ \emph {et~al.}(2003)\citenamefont
  {Leibfried}, \citenamefont {DeMarco}, \citenamefont {Meyer}, \citenamefont
  {Lucas}, \citenamefont {Barrett}, \citenamefont {Britton}, \citenamefont
  {Itano}, \citenamefont {Jelenkovi{\' c}}, \citenamefont {Langer},
  \citenamefont {Rosenband},\ and\ \citenamefont {Wineland}}]{leibfried03}%
  \BibitemOpen
  \bibfield  {author} {\bibinfo {author} {\bibfnamefont {D.}~\bibnamefont
  {Leibfried}}, \bibinfo {author} {\bibfnamefont {B.}~\bibnamefont {DeMarco}},
  \bibinfo {author} {\bibfnamefont {V.}~\bibnamefont {Meyer}}, \bibinfo
  {author} {\bibfnamefont {D.}~\bibnamefont {Lucas}}, \bibinfo {author}
  {\bibfnamefont {M.}~\bibnamefont {Barrett}}, \bibinfo {author} {\bibfnamefont
  {J.}~\bibnamefont {Britton}}, \bibinfo {author} {\bibfnamefont {W.~M.}\
  \bibnamefont {Itano}}, \bibinfo {author} {\bibfnamefont {B.}~\bibnamefont
  {Jelenkovi{\' c}}}, \bibinfo {author} {\bibfnamefont {C.}~\bibnamefont
  {Langer}}, \bibinfo {author} {\bibfnamefont {T.}~\bibnamefont {Rosenband}}, \
  and\ \bibinfo {author} {\bibfnamefont {D.~J.}\ \bibnamefont {Wineland}},\
  }\href {\doibase 10.1038/nature01492} {\bibfield  {journal} {\bibinfo
  {journal} {Nature}\ }\textbf {\bibinfo {volume} {422}},\ \bibinfo {pages}
  {412} (\bibinfo {year} {2003})}\BibitemShut {NoStop}%
\bibitem [{\citenamefont {Lee}\ \emph {et~al.}(2005)\citenamefont {Lee},
  \citenamefont {Brickman}, \citenamefont {Deslauriers}, \citenamefont
  {Haljan}, \citenamefont {Duan},\ and\ \citenamefont {Monroe}}]{lee05}%
  \BibitemOpen
  \bibfield  {author} {\bibinfo {author} {\bibfnamefont {P.~J.}\ \bibnamefont
  {Lee}}, \bibinfo {author} {\bibfnamefont {K.-A.}\ \bibnamefont {Brickman}},
  \bibinfo {author} {\bibfnamefont {L.}~\bibnamefont {Deslauriers}}, \bibinfo
  {author} {\bibfnamefont {P.~C.}\ \bibnamefont {Haljan}}, \bibinfo {author}
  {\bibfnamefont {L.-M.}\ \bibnamefont {Duan}}, \ and\ \bibinfo {author}
  {\bibfnamefont {C.}~\bibnamefont {Monroe}},\ }\href {\doibase
  10.1088/1464-4266/7/10/025} {\bibfield  {journal} {\bibinfo  {journal} {J.\
  Opt.\ B: Quantum Semiclass.\ Opt.}\ }\textbf {\bibinfo {volume} {7}},\
  \bibinfo {pages} {S371} (\bibinfo {year} {2005})}\BibitemShut {NoStop}%
\bibitem [{\citenamefont {Blais}\ \emph {et~al.}(2004)\citenamefont {Blais},
  \citenamefont {Huang}, \citenamefont {Wallraff}, \citenamefont {Girvin},\
  and\ \citenamefont {Schoelkopf}}]{blais04}%
  \BibitemOpen
  \bibfield  {author} {\bibinfo {author} {\bibfnamefont {A.}~\bibnamefont
  {Blais}}, \bibinfo {author} {\bibfnamefont {R.-S.}\ \bibnamefont {Huang}},
  \bibinfo {author} {\bibfnamefont {A.}~\bibnamefont {Wallraff}}, \bibinfo
  {author} {\bibfnamefont {S.~M.}\ \bibnamefont {Girvin}}, \ and\ \bibinfo
  {author} {\bibfnamefont {R.~J.}\ \bibnamefont {Schoelkopf}},\ }\href
  {\doibase 10.1103/physreva.69.062320} {\bibfield  {journal} {\bibinfo
  {journal} {Phys.\ Rev.\ A}\ }\textbf {\bibinfo {volume} {69}},\ \bibinfo
  {pages} {062320} (\bibinfo {year} {2004})}\BibitemShut {NoStop}%
\bibitem [{\citenamefont {Pedernales}\ \emph {et~al.}(2015)\citenamefont
  {Pedernales}, \citenamefont {Lizuain}, \citenamefont {Felicetti},
  \citenamefont {Romero}, \citenamefont {Lamata},\ and\ \citenamefont
  {Solano}}]{pedernales15}%
  \BibitemOpen
  \bibfield  {author} {\bibinfo {author} {\bibfnamefont {J.~S.}\ \bibnamefont
  {Pedernales}}, \bibinfo {author} {\bibfnamefont {I.}~\bibnamefont {Lizuain}},
  \bibinfo {author} {\bibfnamefont {S.}~\bibnamefont {Felicetti}}, \bibinfo
  {author} {\bibfnamefont {G.}~\bibnamefont {Romero}}, \bibinfo {author}
  {\bibfnamefont {L.}~\bibnamefont {Lamata}}, \ and\ \bibinfo {author}
  {\bibfnamefont {E.}~\bibnamefont {Solano}},\ }\href {\doibase
  10.1038/srep15472} {\bibfield  {journal} {\bibinfo  {journal} {Sci.\ Rep.}\
  }\textbf {\bibinfo {volume} {5}},\ \bibinfo {pages} {15472} (\bibinfo {year}
  {2015})}\BibitemShut {NoStop}%
\bibitem [{\citenamefont {Marshall}\ and\ \citenamefont
  {James}(2016)}]{marshall16}%
  \BibitemOpen
  \bibfield  {author} {\bibinfo {author} {\bibfnamefont {K.}~\bibnamefont
  {Marshall}}\ and\ \bibinfo {author} {\bibfnamefont {D.~F.~V.}\ \bibnamefont
  {James}},\ }\href {\doibase 10.1007/s00340-016-6601-y} {\bibfield  {journal}
  {\bibinfo  {journal} {Appl.\ Phys.\ B}\ }\textbf {\bibinfo {volume} {123}},\
  \bibinfo {pages} {26} (\bibinfo {year} {2016})}\BibitemShut {NoStop}%
\bibitem [{\citenamefont {Blais}\ \emph {et~al.}(2020)\citenamefont {Blais},
  \citenamefont {Girvin},\ and\ \citenamefont {Oliver}}]{blais20}%
  \BibitemOpen
  \bibfield  {author} {\bibinfo {author} {\bibfnamefont {A.}~\bibnamefont
  {Blais}}, \bibinfo {author} {\bibfnamefont {S.~M.}\ \bibnamefont {Girvin}}, \
  and\ \bibinfo {author} {\bibfnamefont {W.~D.}\ \bibnamefont {Oliver}},\
  }\href {\doibase 10.1038/s41567-020-0806-z} {\bibfield  {journal} {\bibinfo
  {journal} {Nat.\ Phys.}\ }\textbf {\bibinfo {volume} {16}},\ \bibinfo {pages}
  {247} (\bibinfo {year} {2020})}\BibitemShut {NoStop}%
\bibitem [{\citenamefont {Gao}\ \emph {et~al.}(2018)\citenamefont {Gao},
  \citenamefont {Lester}, \citenamefont {Zhang}, \citenamefont {Wang},
  \citenamefont {Rosenblum}, \citenamefont {Frunzio}, \citenamefont {Jiang},
  \citenamefont {Girvin},\ and\ \citenamefont {Schoelkopf}}]{gao18}%
  \BibitemOpen
  \bibfield  {author} {\bibinfo {author} {\bibfnamefont {Y.~Y.}\ \bibnamefont
  {Gao}}, \bibinfo {author} {\bibfnamefont {B.~J.}\ \bibnamefont {Lester}},
  \bibinfo {author} {\bibfnamefont {Y.}~\bibnamefont {Zhang}}, \bibinfo
  {author} {\bibfnamefont {C.}~\bibnamefont {Wang}}, \bibinfo {author}
  {\bibfnamefont {S.}~\bibnamefont {Rosenblum}}, \bibinfo {author}
  {\bibfnamefont {L.}~\bibnamefont {Frunzio}}, \bibinfo {author} {\bibfnamefont
  {L.}~\bibnamefont {Jiang}}, \bibinfo {author} {\bibfnamefont {S.~M.}\
  \bibnamefont {Girvin}}, \ and\ \bibinfo {author} {\bibfnamefont {R.~J.}\
  \bibnamefont {Schoelkopf}},\ }\href {\doibase 10.1103/physrevx.8.021073}
  {\bibfield  {journal} {\bibinfo  {journal} {Phys.\ Rev.\ X}\ }\textbf
  {\bibinfo {volume} {8}},\ \bibinfo {pages} {021073} (\bibinfo {year}
  {2018})}\BibitemShut {NoStop}%
\bibitem [{\citenamefont {Sundaresan}\ \emph {et~al.}(2015)\citenamefont
  {Sundaresan}, \citenamefont {Liu}, \citenamefont {Sadri}, \citenamefont
  {Sz{\H o}cs}, \citenamefont {Underwood}, \citenamefont {Malekakhlagh},
  \citenamefont {T{\" u}reci},\ and\ \citenamefont {Houck}}]{sundaresan15}%
  \BibitemOpen
  \bibfield  {author} {\bibinfo {author} {\bibfnamefont {N.~M.}\ \bibnamefont
  {Sundaresan}}, \bibinfo {author} {\bibfnamefont {Y.}~\bibnamefont {Liu}},
  \bibinfo {author} {\bibfnamefont {D.}~\bibnamefont {Sadri}}, \bibinfo
  {author} {\bibfnamefont {L.~J.}\ \bibnamefont {Sz{\H o}cs}}, \bibinfo
  {author} {\bibfnamefont {D.~L.}\ \bibnamefont {Underwood}}, \bibinfo {author}
  {\bibfnamefont {M.}~\bibnamefont {Malekakhlagh}}, \bibinfo {author}
  {\bibfnamefont {H.~E.}\ \bibnamefont {T{\" u}reci}}, \ and\ \bibinfo {author}
  {\bibfnamefont {A.~A.}\ \bibnamefont {Houck}},\ }\href {\doibase
  10.1103/physrevx.5.021035} {\bibfield  {journal} {\bibinfo  {journal} {Phys.
  Rev. X}\ }\textbf {\bibinfo {volume} {5}},\ \bibinfo {pages} {021035}
  (\bibinfo {year} {2015})}\BibitemShut {NoStop}%
\bibitem [{\citenamefont {Koll{\' a}r}\ \emph {et~al.}(2019)\citenamefont
  {Koll{\' a}r}, \citenamefont {Fitzpatrick},\ and\ \citenamefont
  {Houck}}]{kollar19}%
  \BibitemOpen
  \bibfield  {author} {\bibinfo {author} {\bibfnamefont {A.~J.}\ \bibnamefont
  {Koll{\' a}r}}, \bibinfo {author} {\bibfnamefont {M.}~\bibnamefont
  {Fitzpatrick}}, \ and\ \bibinfo {author} {\bibfnamefont {A.~A.}\ \bibnamefont
  {Houck}},\ }\href {\doibase 10.1038/s41586-019-1348-3} {\bibfield  {journal}
  {\bibinfo  {journal} {Nature}\ }\textbf {\bibinfo {volume} {571}},\ \bibinfo
  {pages} {45} (\bibinfo {year} {2019})}\BibitemShut {NoStop}%
\bibitem [{\citenamefont {Gerritsma}\ \emph {et~al.}(2011)\citenamefont
  {Gerritsma}, \citenamefont {Lanyon}, \citenamefont {Kirchmair}, \citenamefont
  {Z{\" a}hringer}, \citenamefont {Hempel}, \citenamefont {Casanova},
  \citenamefont {Garc{\' i}a-Ripoll}, \citenamefont {Solano}, \citenamefont
  {Blatt},\ and\ \citenamefont {Roos}}]{gerritsma11}%
  \BibitemOpen
  \bibfield  {author} {\bibinfo {author} {\bibfnamefont {R.}~\bibnamefont
  {Gerritsma}}, \bibinfo {author} {\bibfnamefont {B.~P.}\ \bibnamefont
  {Lanyon}}, \bibinfo {author} {\bibfnamefont {G.}~\bibnamefont {Kirchmair}},
  \bibinfo {author} {\bibfnamefont {F.}~\bibnamefont {Z{\" a}hringer}},
  \bibinfo {author} {\bibfnamefont {C.}~\bibnamefont {Hempel}}, \bibinfo
  {author} {\bibfnamefont {J.}~\bibnamefont {Casanova}}, \bibinfo {author}
  {\bibfnamefont {J.~J.}\ \bibnamefont {Garc{\' i}a-Ripoll}}, \bibinfo {author}
  {\bibfnamefont {E.}~\bibnamefont {Solano}}, \bibinfo {author} {\bibfnamefont
  {R.}~\bibnamefont {Blatt}}, \ and\ \bibinfo {author} {\bibfnamefont {C.~F.}\
  \bibnamefont {Roos}},\ }\href {\doibase 10.1103/physrevlett.106.060503}
  {\bibfield  {journal} {\bibinfo  {journal} {Phys.\ Rev.\ Lett.}\ }\textbf
  {\bibinfo {volume} {106}},\ \bibinfo {pages} {060503} (\bibinfo {year}
  {2011})}\BibitemShut {NoStop}%
\bibitem [{\citenamefont {K{\" u}hl}\ and\ \citenamefont
  {Domcke}(2002)}]{kuhl02}%
  \BibitemOpen
  \bibfield  {author} {\bibinfo {author} {\bibfnamefont {A.}~\bibnamefont {K{\"
  u}hl}}\ and\ \bibinfo {author} {\bibfnamefont {W.}~\bibnamefont {Domcke}},\
  }\href {\doibase 10.1063/1.1423326} {\bibfield  {journal} {\bibinfo
  {journal} {J.\ Chem.\ Phys.}\ }\textbf {\bibinfo {volume} {116}},\ \bibinfo
  {pages} {263} (\bibinfo {year} {2002})}\BibitemShut {NoStop}%
\bibitem [{\citenamefont {Barreiro}\ \emph {et~al.}(2011)\citenamefont
  {Barreiro}, \citenamefont {M{\" u}ller}, \citenamefont {Schindler},
  \citenamefont {Nigg}, \citenamefont {Monz}, \citenamefont {Chwalla},
  \citenamefont {Hennrich}, \citenamefont {Roos}, \citenamefont {Zoller},\ and\
  \citenamefont {Blatt}}]{barreiro11}%
  \BibitemOpen
  \bibfield  {author} {\bibinfo {author} {\bibfnamefont {J.~T.}\ \bibnamefont
  {Barreiro}}, \bibinfo {author} {\bibfnamefont {M.}~\bibnamefont {M{\"
  u}ller}}, \bibinfo {author} {\bibfnamefont {P.}~\bibnamefont {Schindler}},
  \bibinfo {author} {\bibfnamefont {D.}~\bibnamefont {Nigg}}, \bibinfo {author}
  {\bibfnamefont {T.}~\bibnamefont {Monz}}, \bibinfo {author} {\bibfnamefont
  {M.}~\bibnamefont {Chwalla}}, \bibinfo {author} {\bibfnamefont
  {M.}~\bibnamefont {Hennrich}}, \bibinfo {author} {\bibfnamefont {C.~F.}\
  \bibnamefont {Roos}}, \bibinfo {author} {\bibfnamefont {P.}~\bibnamefont
  {Zoller}}, \ and\ \bibinfo {author} {\bibfnamefont {R.}~\bibnamefont
  {Blatt}},\ }\href {\doibase 10.1038/nature09801} {\bibfield  {journal}
  {\bibinfo  {journal} {Nature}\ }\textbf {\bibinfo {volume} {470}},\ \bibinfo
  {pages} {486} (\bibinfo {year} {2011})}\BibitemShut {NoStop}%
\bibitem [{\citenamefont {Nitzan}(2006)}]{Nitzan2006}%
  \BibitemOpen
  \bibfield  {author} {\bibinfo {author} {\bibfnamefont {A.}~\bibnamefont
  {Nitzan}},\ }\href@noop {} {\emph {\bibinfo {title} {{Chemical Dynamics in
  Condensed Phases}}}}\ (\bibinfo  {publisher} {Oxford University Press},\
  \bibinfo {year} {2006})\BibitemShut {NoStop}%
\bibitem [{\citenamefont {Stenholm}(1986)}]{stenholm86}%
  \BibitemOpen
  \bibfield  {author} {\bibinfo {author} {\bibfnamefont {S.}~\bibnamefont
  {Stenholm}},\ }\href {\doibase 10.1103/revmodphys.58.699} {\bibfield
  {journal} {\bibinfo  {journal} {Rev.\ Mod.\ Phys.}\ }\textbf {\bibinfo
  {volume} {58}},\ \bibinfo {pages} {699} (\bibinfo {year} {1986})}\BibitemShut
  {NoStop}%
\bibitem [{\citenamefont {Diedrich}\ \emph {et~al.}(1989)\citenamefont
  {Diedrich}, \citenamefont {Bergquist}, \citenamefont {Itano},\ and\
  \citenamefont {Wineland}}]{Diedrich.1989}%
  \BibitemOpen
  \bibfield  {author} {\bibinfo {author} {\bibfnamefont {F.}~\bibnamefont
  {Diedrich}}, \bibinfo {author} {\bibfnamefont {J.~C.}\ \bibnamefont
  {Bergquist}}, \bibinfo {author} {\bibfnamefont {W.~M.}\ \bibnamefont
  {Itano}}, \ and\ \bibinfo {author} {\bibfnamefont {D.~J.}\ \bibnamefont
  {Wineland}},\ }\href@noop {} {\bibfield  {journal} {\bibinfo  {journal}
  {Phys. Rev. Lett.}\ }\textbf {\bibinfo {volume} {62}},\ \bibinfo {pages}
  {403} (\bibinfo {year} {1989})}\BibitemShut {NoStop}%
\bibitem [{\citenamefont {Lechner}\ \emph {et~al.}(2016)\citenamefont
  {Lechner}, \citenamefont {Maier}, \citenamefont {Hempel}, \citenamefont
  {Jurcevic}, \citenamefont {Lanyon}, \citenamefont {Monz}, \citenamefont
  {Brownnutt}, \citenamefont {Blatt},\ and\ \citenamefont
  {Roos}}]{Lechner.2016}%
  \BibitemOpen
  \bibfield  {author} {\bibinfo {author} {\bibfnamefont {R.}~\bibnamefont
  {Lechner}}, \bibinfo {author} {\bibfnamefont {C.}~\bibnamefont {Maier}},
  \bibinfo {author} {\bibfnamefont {C.}~\bibnamefont {Hempel}}, \bibinfo
  {author} {\bibfnamefont {P.}~\bibnamefont {Jurcevic}}, \bibinfo {author}
  {\bibfnamefont {B.~P.}\ \bibnamefont {Lanyon}}, \bibinfo {author}
  {\bibfnamefont {T.}~\bibnamefont {Monz}}, \bibinfo {author} {\bibfnamefont
  {M.}~\bibnamefont {Brownnutt}}, \bibinfo {author} {\bibfnamefont
  {R.}~\bibnamefont {Blatt}}, \ and\ \bibinfo {author} {\bibfnamefont {C.~F.}\
  \bibnamefont {Roos}},\ }\href@noop {} {\bibfield  {journal} {\bibinfo
  {journal} {Phys. Rev. A}\ }\textbf {\bibinfo {volume} {93}},\ \bibinfo
  {pages} {053401} (\bibinfo {year} {2016})}\BibitemShut {NoStop}%
\bibitem [{\citenamefont {Lucas}\ \emph {et~al.}(2007)\citenamefont {Lucas},
  \citenamefont {Keitch}, \citenamefont {Home}, \citenamefont {Imreh},
  \citenamefont {McDonnell}, \citenamefont {Stacey}, \citenamefont {Szwer},\
  and\ \citenamefont {Steane}}]{lucas07}%
  \BibitemOpen
  \bibfield  {author} {\bibinfo {author} {\bibfnamefont {D.~M.}\ \bibnamefont
  {Lucas}}, \bibinfo {author} {\bibfnamefont {B.~C.}\ \bibnamefont {Keitch}},
  \bibinfo {author} {\bibfnamefont {J.~P.}\ \bibnamefont {Home}}, \bibinfo
  {author} {\bibfnamefont {G.}~\bibnamefont {Imreh}}, \bibinfo {author}
  {\bibfnamefont {M.~J.}\ \bibnamefont {McDonnell}}, \bibinfo {author}
  {\bibfnamefont {D.~N.}\ \bibnamefont {Stacey}}, \bibinfo {author}
  {\bibfnamefont {D.~J.}\ \bibnamefont {Szwer}}, \ and\ \bibinfo {author}
  {\bibfnamefont {A.~M.}\ \bibnamefont {Steane}},\ }\href@noop {} {\bibfield
  {journal} {\bibinfo  {journal} {arXiv}\ } (\bibinfo {year} {2007})},\ \Eprint
  {http://arxiv.org/abs/0710.4421} {0710.4421} \BibitemShut {NoStop}%
\bibitem [{\citenamefont {Talukdar}\ \emph {et~al.}(2016)\citenamefont
  {Talukdar}, \citenamefont {Gorman}, \citenamefont {Daniilidis}, \citenamefont
  {Schindler}, \citenamefont {Ebadi}, \citenamefont {Kaufmann}, \citenamefont
  {Zhang},\ and\ \citenamefont {H{\" a}ffner}}]{talukdar16}%
  \BibitemOpen
  \bibfield  {author} {\bibinfo {author} {\bibfnamefont {I.}~\bibnamefont
  {Talukdar}}, \bibinfo {author} {\bibfnamefont {D.~J.}\ \bibnamefont
  {Gorman}}, \bibinfo {author} {\bibfnamefont {N.}~\bibnamefont {Daniilidis}},
  \bibinfo {author} {\bibfnamefont {P.}~\bibnamefont {Schindler}}, \bibinfo
  {author} {\bibfnamefont {S.}~\bibnamefont {Ebadi}}, \bibinfo {author}
  {\bibfnamefont {H.}~\bibnamefont {Kaufmann}}, \bibinfo {author}
  {\bibfnamefont {T.}~\bibnamefont {Zhang}}, \ and\ \bibinfo {author}
  {\bibfnamefont {H.}~\bibnamefont {H{\" a}ffner}},\ }\href {\doibase
  10.1103/PhysRevA.93.043415} {\bibfield  {journal} {\bibinfo  {journal}
  {Phys.\ Rev.\ A}\ }\textbf {\bibinfo {volume} {93}},\ \bibinfo {pages}
  {043415} (\bibinfo {year} {2016})}\BibitemShut {NoStop}%
\bibitem [{\citenamefont {Wang}\ \emph {et~al.}(2009)\citenamefont {Wang},
  \citenamefont {Hofheinz}, \citenamefont {Ansmann}, \citenamefont {Bialczak},
  \citenamefont {Lucero}, \citenamefont {Neeley}, \citenamefont {O'Connell},
  \citenamefont {Sank}, \citenamefont {Weides}, \citenamefont {Wenner},
  \citenamefont {Cleland},\ and\ \citenamefont {Martinis}}]{wang09}%
  \BibitemOpen
  \bibfield  {author} {\bibinfo {author} {\bibfnamefont {H.}~\bibnamefont
  {Wang}}, \bibinfo {author} {\bibfnamefont {M.}~\bibnamefont {Hofheinz}},
  \bibinfo {author} {\bibfnamefont {M.}~\bibnamefont {Ansmann}}, \bibinfo
  {author} {\bibfnamefont {R.~C.}\ \bibnamefont {Bialczak}}, \bibinfo {author}
  {\bibfnamefont {E.}~\bibnamefont {Lucero}}, \bibinfo {author} {\bibfnamefont
  {M.}~\bibnamefont {Neeley}}, \bibinfo {author} {\bibfnamefont {A.~D.}\
  \bibnamefont {O'Connell}}, \bibinfo {author} {\bibfnamefont {D.}~\bibnamefont
  {Sank}}, \bibinfo {author} {\bibfnamefont {M.}~\bibnamefont {Weides}},
  \bibinfo {author} {\bibfnamefont {J.}~\bibnamefont {Wenner}}, \bibinfo
  {author} {\bibfnamefont {A.~N.}\ \bibnamefont {Cleland}}, \ and\ \bibinfo
  {author} {\bibfnamefont {J.~M.}\ \bibnamefont {Martinis}},\ }\href {\doibase
  10.1103/PhysRevLett.103.200404} {\bibfield  {journal} {\bibinfo  {journal}
  {Phys.\ Rev.\ Lett.}\ }\textbf {\bibinfo {volume} {103}},\ \bibinfo {pages}
  {200404} (\bibinfo {year} {2009})}\BibitemShut {NoStop}%
\bibitem [{\citenamefont {Sarovar}\ \emph {et~al.}(2017)\citenamefont
  {Sarovar}, \citenamefont {Zhang},\ and\ \citenamefont {Zeng}}]{sarovar17}%
  \BibitemOpen
  \bibfield  {author} {\bibinfo {author} {\bibfnamefont {M.}~\bibnamefont
  {Sarovar}}, \bibinfo {author} {\bibfnamefont {J.}~\bibnamefont {Zhang}}, \
  and\ \bibinfo {author} {\bibfnamefont {L.}~\bibnamefont {Zeng}},\ }\href
  {\doibase 10.1140/epjqt/s40507-016-0054-4} {\bibfield  {journal} {\bibinfo
  {journal} {EPJ Quantum Technol.}\ }\textbf {\bibinfo {volume} {4}},\ \bibinfo
  {pages} {1} (\bibinfo {year} {2017})}\BibitemShut {NoStop}%
\bibitem [{\citenamefont {Poggi}\ \emph {et~al.}(2020)\citenamefont {Poggi},
  \citenamefont {Lysne}, \citenamefont {Kuper}, \citenamefont {Deutsch},\ and\
  \citenamefont {Jessen}}]{poggi20}%
  \BibitemOpen
  \bibfield  {author} {\bibinfo {author} {\bibfnamefont {P.~M.}\ \bibnamefont
  {Poggi}}, \bibinfo {author} {\bibfnamefont {N.~K.}\ \bibnamefont {Lysne}},
  \bibinfo {author} {\bibfnamefont {K.~W.}\ \bibnamefont {Kuper}}, \bibinfo
  {author} {\bibfnamefont {I.~H.}\ \bibnamefont {Deutsch}}, \ and\ \bibinfo
  {author} {\bibfnamefont {P.~S.}\ \bibnamefont {Jessen}},\ }\href {\doibase
  10.1103/PRXQuantum.1.020308} {\bibfield  {journal} {\bibinfo  {journal} {PRX
  Quantum}\ }\textbf {\bibinfo {volume} {1}},\ \bibinfo {pages} {020308}
  (\bibinfo {year} {2020})}\BibitemShut {NoStop}%
\bibitem [{\citenamefont {Friis}\ \emph {et~al.}(2018)\citenamefont {Friis},
  \citenamefont {Marty}, \citenamefont {Maier}, \citenamefont {Hempel},
  \citenamefont {Holz{\" a}pfel}, \citenamefont {Jurcevic}, \citenamefont
  {Plenio}, \citenamefont {Huber}, \citenamefont {Roos}, \citenamefont
  {Blatt},\ and\ \citenamefont {Lanyon}}]{friis18}%
  \BibitemOpen
  \bibfield  {author} {\bibinfo {author} {\bibfnamefont {N.}~\bibnamefont
  {Friis}}, \bibinfo {author} {\bibfnamefont {O.}~\bibnamefont {Marty}},
  \bibinfo {author} {\bibfnamefont {C.}~\bibnamefont {Maier}}, \bibinfo
  {author} {\bibfnamefont {C.}~\bibnamefont {Hempel}}, \bibinfo {author}
  {\bibfnamefont {M.}~\bibnamefont {Holz{\" a}pfel}}, \bibinfo {author}
  {\bibfnamefont {P.}~\bibnamefont {Jurcevic}}, \bibinfo {author}
  {\bibfnamefont {M.~B.}\ \bibnamefont {Plenio}}, \bibinfo {author}
  {\bibfnamefont {M.}~\bibnamefont {Huber}}, \bibinfo {author} {\bibfnamefont
  {C.}~\bibnamefont {Roos}}, \bibinfo {author} {\bibfnamefont {R.}~\bibnamefont
  {Blatt}}, \ and\ \bibinfo {author} {\bibfnamefont {B.}~\bibnamefont
  {Lanyon}},\ }\href {\doibase 10.1103/PhysRevX.8.021012} {\bibfield  {journal}
  {\bibinfo  {journal} {Phys.\ Rev.\ X}\ }\textbf {\bibinfo {volume} {8}},\
  \bibinfo {pages} {021012} (\bibinfo {year} {2018})}\BibitemShut {NoStop}%
\bibitem [{\citenamefont {S{\o}rensen}\ and\ \citenamefont
  {M{\o}lmer}(1999)}]{sorensen99}%
  \BibitemOpen
  \bibfield  {author} {\bibinfo {author} {\bibfnamefont {A.}~\bibnamefont
  {S{\o}rensen}}\ and\ \bibinfo {author} {\bibfnamefont {K.}~\bibnamefont
  {M{\o}lmer}},\ }\href {\doibase 10.1103/PhysRevLett.82.1971} {\bibfield
  {journal} {\bibinfo  {journal} {Phys.\ Rev.\ Lett.}\ }\textbf {\bibinfo
  {volume} {82}},\ \bibinfo {pages} {1971} (\bibinfo {year}
  {1999})}\BibitemShut {NoStop}%
\bibitem [{\citenamefont {Hempel}\ \emph {et~al.}(2013)\citenamefont {Hempel},
  \citenamefont {Lanyon}, \citenamefont {Jurcevic}, \citenamefont {Gerritsma},
  \citenamefont {Blatt},\ and\ \citenamefont {Roos}}]{Hempel.2013}%
  \BibitemOpen
  \bibfield  {author} {\bibinfo {author} {\bibfnamefont {C.}~\bibnamefont
  {Hempel}}, \bibinfo {author} {\bibfnamefont {B.~P.}\ \bibnamefont {Lanyon}},
  \bibinfo {author} {\bibfnamefont {P.}~\bibnamefont {Jurcevic}}, \bibinfo
  {author} {\bibfnamefont {R.}~\bibnamefont {Gerritsma}}, \bibinfo {author}
  {\bibfnamefont {R.}~\bibnamefont {Blatt}}, \ and\ \bibinfo {author}
  {\bibfnamefont {C.~F.}\ \bibnamefont {Roos}},\ }\href {\doibase
  10.1038/nphoton.2013.172} {\bibfield  {journal} {\bibinfo  {journal} {Nature
  Photon.}\ }\textbf {\bibinfo {volume} {7}},\ \bibinfo {pages} {630} (\bibinfo
  {year} {2013})}\BibitemShut {NoStop}%
\bibitem [{\citenamefont {Hosaka}\ \emph {et~al.}(2009)\citenamefont {Hosaka},
  \citenamefont {Webster}, \citenamefont {Stannard}, \citenamefont {Walton},
  \citenamefont {Margolis},\ and\ \citenamefont {Gill}}]{Hosaka.2009}%
  \BibitemOpen
  \bibfield  {author} {\bibinfo {author} {\bibfnamefont {K.}~\bibnamefont
  {Hosaka}}, \bibinfo {author} {\bibfnamefont {S.~A.}\ \bibnamefont {Webster}},
  \bibinfo {author} {\bibfnamefont {A.}~\bibnamefont {Stannard}}, \bibinfo
  {author} {\bibfnamefont {B.~R.}\ \bibnamefont {Walton}}, \bibinfo {author}
  {\bibfnamefont {H.~S.}\ \bibnamefont {Margolis}}, \ and\ \bibinfo {author}
  {\bibfnamefont {P.}~\bibnamefont {Gill}},\ }\href@noop {} {\bibfield
  {journal} {\bibinfo  {journal} {Phys. Rev. A}\ }\textbf {\bibinfo {volume}
  {79}},\ \bibinfo {pages} {033403} (\bibinfo {year} {2009})}\BibitemShut
  {NoStop}%
\bibitem [{\citenamefont {Edmunds}\ \emph {et~al.}(2020)\citenamefont
  {Edmunds}, \citenamefont {Tan}, \citenamefont {Milne}, \citenamefont {Singh},
  \citenamefont {Biercuk},\ and\ \citenamefont {Hempel}}]{Edmunds.2020b}%
  \BibitemOpen
  \bibfield  {author} {\bibinfo {author} {\bibfnamefont {C.~L.}\ \bibnamefont
  {Edmunds}}, \bibinfo {author} {\bibfnamefont {T.~R.}\ \bibnamefont {Tan}},
  \bibinfo {author} {\bibfnamefont {A.}~\bibnamefont {Milne}}, \bibinfo
  {author} {\bibfnamefont {A.}~\bibnamefont {Singh}}, \bibinfo {author}
  {\bibfnamefont {M.}~\bibnamefont {Biercuk}}, \ and\ \bibinfo {author}
  {\bibfnamefont {C.}~\bibnamefont {Hempel}},\ }\href@noop {} {\bibfield
  {journal} {\bibinfo  {journal} {in preparation}\ } (\bibinfo {year}
  {2020})}\BibitemShut {NoStop}%
\bibitem [{\citenamefont {Vlastakis}\ \emph {et~al.}(2013)\citenamefont
  {Vlastakis}, \citenamefont {Kirchmair}, \citenamefont {Leghtas},
  \citenamefont {Nigg}, \citenamefont {Frunzio}, \citenamefont {Girvin},
  \citenamefont {Mirrahimi}, \citenamefont {Devoret},\ and\ \citenamefont
  {Schoelkopf}}]{Vlastakis.2013}%
  \BibitemOpen
  \bibfield  {author} {\bibinfo {author} {\bibfnamefont {B.}~\bibnamefont
  {Vlastakis}}, \bibinfo {author} {\bibfnamefont {G.}~\bibnamefont
  {Kirchmair}}, \bibinfo {author} {\bibfnamefont {Z.}~\bibnamefont {Leghtas}},
  \bibinfo {author} {\bibfnamefont {S.~E.}\ \bibnamefont {Nigg}}, \bibinfo
  {author} {\bibfnamefont {L.}~\bibnamefont {Frunzio}}, \bibinfo {author}
  {\bibfnamefont {S.~M.}\ \bibnamefont {Girvin}}, \bibinfo {author}
  {\bibfnamefont {M.}~\bibnamefont {Mirrahimi}}, \bibinfo {author}
  {\bibfnamefont {M.~H.}\ \bibnamefont {Devoret}}, \ and\ \bibinfo {author}
  {\bibfnamefont {R.~J.}\ \bibnamefont {Schoelkopf}},\ }\href {\doibase
  10.1126/science.1243289} {\bibfield  {journal} {\bibinfo  {journal}
  {Science}\ }\textbf {\bibinfo {volume} {342}},\ \bibinfo {pages} {607}
  (\bibinfo {year} {2013})}\BibitemShut {NoStop}%
\bibitem [{\citenamefont {Harris}\ \emph {et~al.}(2020)\citenamefont {Harris},
  \citenamefont {Millman}, \citenamefont {van~der Walt}, \citenamefont
  {Gommers}, \citenamefont {Virtanen}, \citenamefont {Cournapeau},
  \citenamefont {Wieser}, \citenamefont {Taylor}, \citenamefont {Berg},
  \citenamefont {Smith}, \citenamefont {Kern}, \citenamefont {Picus},
  \citenamefont {Hoyer}, \citenamefont {van Kerkwijk}, \citenamefont {Brett},
  \citenamefont {Haldane}, \citenamefont {del R{\' i}o}, \citenamefont {Wiebe},
  \citenamefont {Peterson}, \citenamefont {G{\' e}rard-Marchant}, \citenamefont
  {Sheppard}, \citenamefont {Reddy}, \citenamefont {Weckesser}, \citenamefont
  {Abbasi}, \citenamefont {Gohlke},\ and\ \citenamefont {Oliphant}}]{numpy}%
  \BibitemOpen
  \bibfield  {author} {\bibinfo {author} {\bibfnamefont {C.~R.}\ \bibnamefont
  {Harris}}, \bibinfo {author} {\bibfnamefont {K.~J.}\ \bibnamefont {Millman}},
  \bibinfo {author} {\bibfnamefont {S.~J.}\ \bibnamefont {van~der Walt}},
  \bibinfo {author} {\bibfnamefont {R.}~\bibnamefont {Gommers}}, \bibinfo
  {author} {\bibfnamefont {P.}~\bibnamefont {Virtanen}}, \bibinfo {author}
  {\bibfnamefont {D.}~\bibnamefont {Cournapeau}}, \bibinfo {author}
  {\bibfnamefont {E.}~\bibnamefont {Wieser}}, \bibinfo {author} {\bibfnamefont
  {J.}~\bibnamefont {Taylor}}, \bibinfo {author} {\bibfnamefont
  {S.}~\bibnamefont {Berg}}, \bibinfo {author} {\bibfnamefont {N.~J.}\
  \bibnamefont {Smith}}, \bibinfo {author} {\bibfnamefont {R.}~\bibnamefont
  {Kern}}, \bibinfo {author} {\bibfnamefont {M.}~\bibnamefont {Picus}},
  \bibinfo {author} {\bibfnamefont {S.}~\bibnamefont {Hoyer}}, \bibinfo
  {author} {\bibfnamefont {M.~H.}\ \bibnamefont {van Kerkwijk}}, \bibinfo
  {author} {\bibfnamefont {M.}~\bibnamefont {Brett}}, \bibinfo {author}
  {\bibfnamefont {A.}~\bibnamefont {Haldane}}, \bibinfo {author} {\bibfnamefont
  {J.~F.}\ \bibnamefont {del R{\' i}o}}, \bibinfo {author} {\bibfnamefont
  {M.}~\bibnamefont {Wiebe}}, \bibinfo {author} {\bibfnamefont
  {P.}~\bibnamefont {Peterson}}, \bibinfo {author} {\bibfnamefont
  {P.}~\bibnamefont {G{\' e}rard-Marchant}}, \bibinfo {author} {\bibfnamefont
  {K.}~\bibnamefont {Sheppard}}, \bibinfo {author} {\bibfnamefont
  {T.}~\bibnamefont {Reddy}}, \bibinfo {author} {\bibfnamefont
  {W.}~\bibnamefont {Weckesser}}, \bibinfo {author} {\bibfnamefont
  {H.}~\bibnamefont {Abbasi}}, \bibinfo {author} {\bibfnamefont
  {C.}~\bibnamefont {Gohlke}}, \ and\ \bibinfo {author} {\bibfnamefont {T.~E.}\
  \bibnamefont {Oliphant}},\ }\href {\doibase 10.1038/s41586-020-2649-2}
  {\bibfield  {journal} {\bibinfo  {journal} {Nature}\ }\textbf {\bibinfo
  {volume} {585}},\ \bibinfo {pages} {357} (\bibinfo {year}
  {2020})}\BibitemShut {NoStop}%
\bibitem [{\citenamefont {Virtanen}\ \emph {et~al.}(2020)\citenamefont
  {Virtanen}, \citenamefont {Gommers}, \citenamefont {Oliphant}, \citenamefont
  {Haberland}, \citenamefont {Reddy}, \citenamefont {Cournapeau}, \citenamefont
  {Burovski}, \citenamefont {Peterson}, \citenamefont {Weckesser},
  \citenamefont {Bright}, \citenamefont {van~der Walt}, \citenamefont {Brett},
  \citenamefont {Wilson}, \citenamefont {Millman}, \citenamefont {Mayorov},
  \citenamefont {Nelson}, \citenamefont {Jones}, \citenamefont {Kern},
  \citenamefont {Larson}, \citenamefont {Carey}, \citenamefont {Polat},
  \citenamefont {Feng}, \citenamefont {Moore}, \citenamefont {VanderPlas},
  \citenamefont {Laxalde}, \citenamefont {Perktold}, \citenamefont {Cimrman},
  \citenamefont {Henriksen}, \citenamefont {Quintero}, \citenamefont {Harris},
  \citenamefont {Archibald}, \citenamefont {Ribeiro}, \citenamefont
  {Pedregosa}, \citenamefont {van Mulbregt},\ and\ \citenamefont {{SciPy 1.0
  Contributors}}}]{scipy}%
  \BibitemOpen
  \bibfield  {author} {\bibinfo {author} {\bibfnamefont {P.}~\bibnamefont
  {Virtanen}}, \bibinfo {author} {\bibfnamefont {R.}~\bibnamefont {Gommers}},
  \bibinfo {author} {\bibfnamefont {T.~E.}\ \bibnamefont {Oliphant}}, \bibinfo
  {author} {\bibfnamefont {M.}~\bibnamefont {Haberland}}, \bibinfo {author}
  {\bibfnamefont {T.}~\bibnamefont {Reddy}}, \bibinfo {author} {\bibfnamefont
  {D.}~\bibnamefont {Cournapeau}}, \bibinfo {author} {\bibfnamefont
  {E.}~\bibnamefont {Burovski}}, \bibinfo {author} {\bibfnamefont
  {P.}~\bibnamefont {Peterson}}, \bibinfo {author} {\bibfnamefont
  {W.}~\bibnamefont {Weckesser}}, \bibinfo {author} {\bibfnamefont
  {J.}~\bibnamefont {Bright}}, \bibinfo {author} {\bibfnamefont {S.~J.}\
  \bibnamefont {van~der Walt}}, \bibinfo {author} {\bibfnamefont
  {M.}~\bibnamefont {Brett}}, \bibinfo {author} {\bibfnamefont
  {J.}~\bibnamefont {Wilson}}, \bibinfo {author} {\bibfnamefont {K.~J.}\
  \bibnamefont {Millman}}, \bibinfo {author} {\bibfnamefont {N.}~\bibnamefont
  {Mayorov}}, \bibinfo {author} {\bibfnamefont {A.~R.~J.}\ \bibnamefont
  {Nelson}}, \bibinfo {author} {\bibfnamefont {E.}~\bibnamefont {Jones}},
  \bibinfo {author} {\bibfnamefont {R.}~\bibnamefont {Kern}}, \bibinfo {author}
  {\bibfnamefont {E.}~\bibnamefont {Larson}}, \bibinfo {author} {\bibfnamefont
  {C.~J.}\ \bibnamefont {Carey}}, \bibinfo {author} {\bibfnamefont
  {I.}~\bibnamefont {Polat}}, \bibinfo {author} {\bibfnamefont
  {Y.}~\bibnamefont {Feng}}, \bibinfo {author} {\bibfnamefont {E.~W.}\
  \bibnamefont {Moore}}, \bibinfo {author} {\bibfnamefont {J.}~\bibnamefont
  {VanderPlas}}, \bibinfo {author} {\bibfnamefont {D.}~\bibnamefont {Laxalde}},
  \bibinfo {author} {\bibfnamefont {J.}~\bibnamefont {Perktold}}, \bibinfo
  {author} {\bibfnamefont {R.}~\bibnamefont {Cimrman}}, \bibinfo {author}
  {\bibfnamefont {I.}~\bibnamefont {Henriksen}}, \bibinfo {author}
  {\bibfnamefont {E.~A.}\ \bibnamefont {Quintero}}, \bibinfo {author}
  {\bibfnamefont {C.~R.}\ \bibnamefont {Harris}}, \bibinfo {author}
  {\bibfnamefont {A.~M.}\ \bibnamefont {Archibald}}, \bibinfo {author}
  {\bibfnamefont {A.~H.}\ \bibnamefont {Ribeiro}}, \bibinfo {author}
  {\bibfnamefont {F.}~\bibnamefont {Pedregosa}}, \bibinfo {author}
  {\bibfnamefont {P.}~\bibnamefont {van Mulbregt}}, \ and\ \bibinfo {author}
  {\bibnamefont {{SciPy 1.0 Contributors}}},\ }\href {\doibase
  10.1038/s41592-019-0686-2} {\bibfield  {journal} {\bibinfo  {journal} {Nat.\
  Methods}\ }\textbf {\bibinfo {volume} {17}},\ \bibinfo {pages} {261}
  (\bibinfo {year} {2020})}\BibitemShut {NoStop}%
\bibitem [{\citenamefont {Johansson}\ \emph {et~al.}(2013)\citenamefont
  {Johansson}, \citenamefont {Nation},\ and\ \citenamefont {Nori}}]{qutip}%
  \BibitemOpen
  \bibfield  {author} {\bibinfo {author} {\bibfnamefont {J.~R.}\ \bibnamefont
  {Johansson}}, \bibinfo {author} {\bibfnamefont {P.~D.}\ \bibnamefont
  {Nation}}, \ and\ \bibinfo {author} {\bibfnamefont {F.}~\bibnamefont
  {Nori}},\ }\href {\doibase 10.1016/j.cpc.2012.11.019} {\bibfield  {journal}
  {\bibinfo  {journal} {Comp.\ Phys.\ Comm.}\ }\textbf {\bibinfo {volume}
  {184}},\ \bibinfo {pages} {1234} (\bibinfo {year} {2013})}\BibitemShut
  {NoStop}%
\end{thebibliography}
\end{document}